\documentclass[]{spie}  %
\usepackage[utf8]{inputenc}
\usepackage[numbers]{natbib}

\usepackage{amsmath,amsfonts,amssymb,bm}
\usepackage{graphicx}
\usepackage[colorlinks=true, allcolors=blue]{hyperref}
\usepackage{booktabs}
\usepackage{url,bm,booktabs,multicol,multirow, graphicx}
\usepackage{adjustbox}

\newcommand{\mm}{\(mm\)}
\newcommand{\cm}{\(cm\)~}

\newcommand{\mz}[1]{{\color{black}{#1}}}

\title{Coronary Atherosclerotic Plaque Characterization with Photon-counting CT: 
a Simulation-based Feasibility Study}

\author[a]{Mengzhou Li}
\author[b]{Mingye Wu}
\author[b]{Jed Pack}
\author[b]{Pengwei Wu}
\author[b]{Bruno De Man}
\author[c]{Adam Wang}
\author[c,d]{Koen Nieman}
\author[a,*]{Ge Wang}
\affil[a]{Biomedical Imaging Center, Center for Biotechnology and Interdisciplinary Research, Department of Biomedical Engineering, School of Engineering, Rensselaer Polytechnic Institute, Troy, NY, USA}
\affil[b]{GE HealthCare Technology \& Innovation Center, Niskayuna, NY, USA}
\affil[c]{Department of Radiology, Stanford University, CA, USA}
\affil[d]{Department of Medicine (Cardiovascular Medicine), Stanford University, CA, USA}

\authorinfo{Correspondence: Ge Wang, wangg6@rpi.edu}
    
\begin{document}
\maketitle

\begin{abstract}
Recent development of photon-counting CT (PCCT) brings great opportunities for plaque characterization with much-improved spatial resolution and spectral imaging capability. While existing coronary plaque PCCT imaging results are based on detectors made of CZT or CdTe materials, deep-silicon photon-counting detectors have unique performance characteristics and promise distinct imaging capabilities. In this work, we report a systematic simulation study of a deep-silicon PCCT scanner with a new clinically-relevant digital plaque phantom with realistic geometrical parameters and chemical compositions. This work investigates the effects of spatial resolution, noise, motion artifacts, radiation dose, and spectral characterization. Our simulation results suggest that the deep-silicon PCCT design provides adequate spatial resolution for visualizing a necrotic core and quantitation of key plaque features. Advanced denoising techniques and aggressive bowtie filter designs can keep image noise to acceptable levels at this resolution while keeping radiation dose comparable to that of a conventional CT scan. The ultrahigh resolution of PCCT also means an elevated sensitivity to motion artifacts. It is found that  a tolerance of less than 0.4~\mm~residual movement range requires the application of accurate motion correction methods for best plaque imaging quality with PCCT. 
\end{abstract}


\section{Introduction}

Atherosclerosis is a main source of coronary artery diseases, which has a high risk of acute myocardial infarction or even sudden cardiac death as the first manifestation~\cite{members2016executive,si2021vivo,maurovich2014comprehensive}. Most of those acute coronary events result from rupture of an atherosclerotic plaque while two-thirds of ruptured plaques in sudden luminal thrombi events are characterized by ``a necrotic core covered by a thin layer of fibrous cap''~\cite{maurovich2014comprehensive}. Hence, lesions with a characteristic thin cap (cap thickness \(<65\mu m\)) are often considered to be vulnerable to rupture~\cite{virmani2006pathology}. Atherosclerotic plaque characterization has the potential to identify rupture-prone plaques, risk stratify patients, and guide therapeutics. This is critically important because most infarctions occur in individuals not considered at high-risk, without prior symptoms, and involving non-flow limiting CAD undetectable by functional tests. 

Current assessment of an atherosclerotic plaque is supported by a range of invasive and non-invasive imaging techniques. Invasive techniques like intravascular ultrasound (IVUS) and optical-coherence tomography (OCT) provide reliable results at high resolution and identify rupture-prone plaques with large lipid cores but are unsuitable for screening~\cite{follmer2023roadmap}. Coronary CT angiography (CCTA) demonstrated great promise as a noninvasive alternative in large-cohort studies~\cite{follmer2023roadmap}. However, for individual plaque characterization and monitoring, existing CT scanners lack spatial and contrast resolution to robustly detect small changes in plaque burden and provide inadequate tissue discrimination to identify lipid-rich plaque~\cite{leiner2021new,pack2022cardiac}.

The recent development of photon-counting CT (PCCT) brings new opportunities to address this plaque characterization challenge. Unlike conventional CT with energy-integrating detectors (EIDs), PCCT uses photon-counting detectors (PCDs), which directly converts each incoming X-ray photon into an electronic signal with a semiconductor and counts the number of incident photons rather than measuring an overall intensity through indirect conversion via scintillation. PCCT enjoys the first major advantage of much-improved spatial resolution over EID-CT, due to the absence of septa used between neighboring detector cells in contemporary EIDs for crosstalk prevention which confines the geometric efficiency and ultimately prevent the pixel size from further reduction. On the other hand, PCDs allow recording energy information of detected X-ray photons by classifying them into various energy bins. This multi-bin sensing mechanism provides the second major advantage over the state-of-the-art dual-energy CT for tissue quantification by enhancing spectral information (multiple non-overlapping energy bins for PCCT versus two overlapping spectra for dual-energy CT).

Cadmium Zinc Telluride (CZT) and Cadmium Telluride (CdTe) are the most popular sensor materials for PCDs. The first patient scans with a PCCT prototype, based on a CZT detector, were performed by GE HealthCare as early as 2008~\cite{benjaminov2008novel}. Recently, the first FDA-approved whole body PCCT scanner (equipped with two CdTe detectors) was introduced to the market by Siemens in 2021~\cite{voelker2021advanced}. Similar products from other vendors are also under development.
With those CZT or CdTe detector-based prototypes/products, a few experimental studies were performed for performance evaluation of coronary plaque PCCT imaging with relevant phantoms or patient cohorts.
For example, \cite{si2021vivo} proposed k-edge imaging with gold nanoparticles for macrophage burden detection and quantification in atherosclerotic plaques in live rabbits on a clinical PCCT prototype; \cite{rajagopal2021evaluation} compared plaque and stent imaging results with high-resolution PCCT and conventional EID-CT for stenosis assessment in a phantom study; \cite{rotzinger2021performance} did a similar  comparison with a focus on image quality and plaque detectability in a phantom study; \cite{si2022coronary} compared the CCTA quality with a clinical PCCT prototype and an EID dual-layer CT scanner in patients for the first time; \cite{mergen2022first,mergen2022ultra} investigated the effect of the reconstruction kernel, slice thickness, and matrix size on quantitative coronary plaque characterization and image quality in ultrahigh-resolution CCTA of patients with a PCCT scanner; \cite{zsarnoczay2023ultra} validated the improved accuracy with ultrahigh-resolution PCCT over standard-resolution PCCT for stenosis quantification over various heart rates in a dynamic phantom study; and a few PCCT studies were on the impact of different monoenergetic energy levels on volumetric measurements of plaque components~\cite{vattay2023impact}, image quality of CCTA~\cite{sartoretti2023photon}, and fat tissue attenuation index~\cite{mergen2022epicardial}. In summary, those earlier results experimentally demonstrate the benefits of PCCT for CCTA, including higher resolution resulting in less blooming artifacts and better stenosis quantification, but at the cost of increased image noise. 

More recently GE HealthCare developed a new PCCT scanner technology based on silicon-based detectors with an edge-on design. This deep-silicon detector has unique features that differentiate it from its competitors. Specifically, it offers eight energy bins for unprecedented opportunities of tissue characterization, it promises great stability thanks to the nearly perfect purity of silicon, and it has multiple depth segments to minimize pile-up effects. Since deep-silicon PCCT is a very new technology, it has not yet been widely researched and real patient data are not yet widely accessible.

In this article, we report a systematic simulation study of the deep-silicon PCCT prototype design with a clinical relevant digital plaque phantom with realistic geometrical parameters and chemical compositions. We investigate the resolution needed for plaque characterization, optimize the reconstruction parameters in presence of noise, show the tolerance of high-resolution PCCT imaging to residual motion artifacts, and show how radiation dose and image noise at this ultra-high spatial resolution can be comparable to that of an EID-CT scan.

\section{Methodology}
\subsection{Plaque Modeling}
\mz{Identification of the necrotic core has been a challenging task with traditional EID-CT imaging due to insufficient spatial resolution and contrast resolution. With the latest PCCT technology, spatial resolution of about {0.2~\mm} and much-improved tissue characterization~\cite{rajendran2022first,zhan2023comprehensive,danielsson2021photon} becomes feasible, allowing for unprecedented plaque imaging. To assess the PCCT imaging performance for plaque characterization in a well-controlled environment, here we first design
a digital 3D plaque model with clinically meaningful geometry and realistic chemical composition.}
\subsubsection{Plaque Geometry}

\begin{figure}
  \centering
  \includegraphics*[width=\linewidth]{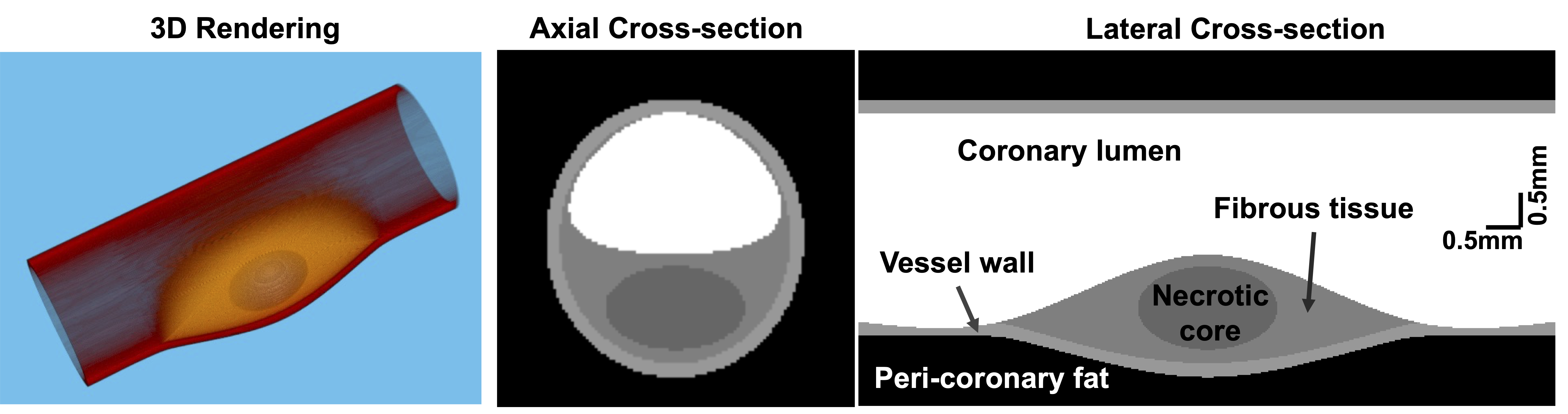}
  \caption{Volumetric rendering of our plaque model, along with central axial section and central sagittal cross-sections illustrating the key plaque components.}\label{fig:plaqueModel}
\end{figure}

We built a digital volumetric non-calcified fibrous plaque phantom for our cardiac CT simulation. Our model consists of five key anatomy structures; i.e., peripheral coronary fat (or peri-coronary fat), coronary vessel wall, lumen filled with blood, fibrous tissue, and a necrotic core, each of which was modeled with a different type of material as illustrated in Fig.~\ref{fig:plaqueModel}.
The original phantom is composed in a binary format with each voxel from one of the materials in an isotropic cube of \(0.021^3~mm^3\) in size. Then, the binary material maps are linearly resized to a smaller volume (with fractional values) of size \(230\times 230 \times 230\) for simulation efficiency with each voxel representing an isotropic cube of \(0.042^3~mm^3\) in size.

\begin{figure}
  \centering
  \includegraphics*[width=0.7\linewidth]{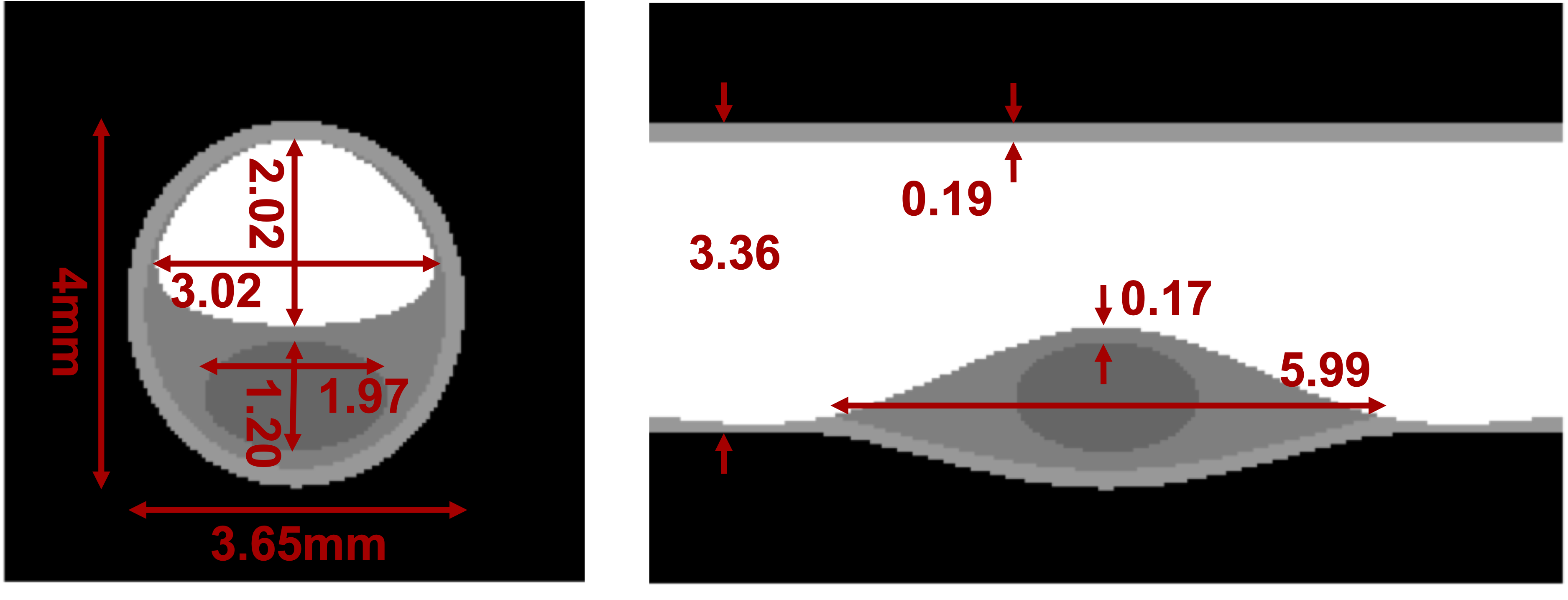}
  \caption{Dimensions for the key features of the plaque components.}\label{fig:plaqueGeom}
\end{figure}

The key structures are geometrically illustrated in Fig.~\ref{fig:plaqueGeom}. The necrotic core is of a flat ellipsoid shape with a major axis length 1.97~\mm~and a minor axis length 1.2~\mm. The diameter of the vessel cross-section is 3.36~\mm~in diameter except for the plaque region where the maximum cross-section is of elliptical shape as shown in the axial cross-section. The vessel wall thickness is set to 0.19~\mm~while the thickness of the fibrous cap is 0.17~\mm. The wall thickness is deviated from the practical range (0.55 to 1.0~\mm)~\cite{fayad2000noninvasive}, while the other geometrical parameters, e.g., lumen size and vessel diameter, are clinically relevant~\cite{perry2013coronary}.

\subsubsection{Chemical Composition}\label{sec:composition}
In reference to the chemical composition model of human coronary artery ~\cite{romer1998histopathology} developed with near-infrared Raman spectroscopy, we use the measured composition of intimal fibroplasia for the vessel wall dry mass, including 62\% delipidized arterial tissue (DA), 6\% free cholesterol (FC), 30\% triglycerides and phospholipids (TG\&PL), and 2\% calcium salt (CS) by relative weight fractions. Then, 30\% dry mass and 70\% water form the final chemical composition. Similarly, the dry mass of fibrous tissue and cap has the composition of 45\% DA, 30\% FC, 15\% cholesterol ester (CE), and 10\% TG\&PL. The water weight ratio in the final composition is 80\%. The necrotic core has more collagen (10\% more) and lipid components (8\% more FC and 8\% more CE) into fibrous tissue and with lower density. The density values for the above materials are tuned to make the corresponding CT values match the targets of 60 Hounsfield  unit (HU), 100 HU, and 40 HU for vessel wall, fibrous tissue, and necrotic core respectively assuming a 100kVp scanning protocol. Those target values are clinically meaningful according to the literature~\cite{maurovich2014comprehensive}. The resultant density values are 1.074\(g/cm^3\), 1.113\(g/cm^3\), and 1.055\(g/cm^3\) respectively.

The chemical composition for peri-coronary fat is set to that of the adipose tissue (ICRU-44), which can be retrieved from the NIST website. The material for the coronary lumen is assumed as blood with 1\% iodine whose composition comes from the NCAT data. The CT values of the two materials are -100 HU and 500 HU respectively for the 100kVp scanning protocol.

More details about the molecular formulas for our model are as follows. The formula for FC is \(\text{C}_{27}\text{H}_{46}\text{O}\)~\cite{Cholesterol}. We use linoleic acid (fatty acid 18:2) based cholesterol ester to represent CE in plaque since it is the most abundant type in human plasma~\cite{nestel1966turnover}, and the corresponding molecular formula is \(\text{C}_{45}\text{H}_{76}\text{O}_{2}\). The phospholipids are omitted due to their small weight fractions (\(<3\%\)). TG is represented by a mixed triglyceride composed of glycerol, palmitic acid (C16:0), oleic acid (C18:1), and \(\alpha\)-linolenic acid (C18:3), since palmitic acid and oleic acid are found most abundant in human lipids~\cite{WATKINS2005186} while \(\alpha\)-linolenic acid is the other essential type fatty acid besides linoleic acid. The resultant formula is \(\text{C}_{55}\text{H}_{98}\text{O}_{6}\). CS is mainly calcium hydroxide phosphate tribasic (hydroxyapatite) with the formula \(\text{HCa}_{5}\text{O}_{13}\text{P}_{3}\). DA is more complicated and mainly composed of different proteins~\cite{buschman2000vivo}, especially from the extracellular matrix which usually contributes over half of the wall mass of arteries or veins~\cite{xu2014vascular}. We represent DA with collagen and elastin while neglecting other complex constituents, like fibrillin and DNA, for simplicity due to their small total weight fractions. The ratio between collagen and elastin is 26\% versus 31\% following the average values measured from aorta walls~\cite{hosoda1984age}. Since artery walls and fibrous plaques consist mostly of types I and III collagen~\cite{shekhonin1987relative} while type I constitutes over 90\% of collagen in the human body, we model collagen simply as type I collagen with a formula of \(\text{C}_{57}\text{H}_{91}\text{N}_{19}\text{O}_{16}\)~\cite{CollagenI}. Similarly, elastin is modeled as the mixture of amino acid residues of the top five abundance in the intimal layer elastin from the normal aortae (the composition changes little for elastin in tunica media with or without mild plaques)~\cite{kramsch1971protein}. The included amino acids and corresponding abundances are Glycine (29.8\%, \(\text{C}_{2}\text{H}_{5}\text{NO}_{2}\)), Alanine (21.1\%, \(\text{C}_{3}\text{H}_{7}\text{NO}_{2}\)), Valine (13.1\%, \(\text{C}_{5}\text{H}_{11}\text{NO}_{2}\)), Proline (12.7\%, \(\text{C}_{5}\text{H}_{9}\text{NO}_{2}\)), and Leucine (6.5\%, \(\text{C}_{6}\text{H}_{13}\text{NO}_{2}\)). Finally, \mz{the basis material compositions are summarized in Table~\ref{table:Explanation}, and the element compositions and density values of the key components are tabulated in Table~\ref{table:composition}}.

\begin{table}[h]
  \renewcommand{\arraystretch}{1.2}
  \caption{\label{table:Explanation}Basis material compositions by weight for plaque modeling.}
  \mz{
  \begin{center}
    \begin{tabular}{lllcccp{7cm}}

      \toprule
      \multicolumn{3}{c}{\multirow{2}{*}{Material Basis}} & Vessel                                  & Fibrous  & Necrotic Core & \multirow{2}{*}{Note}                                                                                                                                                                                                                                                                                                                                                                                                          \\
      \multicolumn{3}{c}{}                                & Wall                                    & Tissue   & (prior norm.) &                                                                                                                                                                                                                                                                                                                                                                                                                                 \\
      \midrule
      \multirow{5}{8truemm}{Dry Mass}                     & \multicolumn{2}{l}{\hspace{3truemm} CE} & -        & 15\%          & 18.3\% (23\%)                                                     & \small{\(\text{C}_{45}\text{H}_{76}\text{O}_{2}\), as a linoleic acid based CE}                                                                                                                                                                                                                                                                             \\
      \cline{2-7}
                                                          & \multicolumn{2}{l}{\hspace{3truemm} FC} & 6\%      & 30\%          & 30.2\% (38\%)                                                     & \small{\(\text{C}_{27}\text{H}_{46}\text{O}\)}, as a free cholesterol                                                                                                                                                                                                                                                                                       \\
      \cline{2-7}
                                                          & \multicolumn{2}{l}{\hspace{3truemm} TG} & 30\%     & 10\%          & 7.9\% (10\%)                                                      & \small{\(\text{C}_{55}\text{H}_{98}\text{O}_{6}\), as a mixed triglyceride (1 glycerol, 1 palmitic acid, 1 oleic acid, 1 \(\alpha\)-linolenic acid)}                                                                                                                                                                                                        \\
      \cline{2-7}
                                                          & \multicolumn{2}{l}{\hspace{3truemm} CS} & 2\%      & -             & -                                                                 & \small{\(\text{HCa}_{5}\text{O}_{13}\text{P}_{3}\), as hydroxyapatite }                                                                                                                                                                                                                                                                                     \\
      \cline{2-7}
                                                          & \multirow{2}{*}{DA}                     & Collagen & 28.3\%        & 20.5\%                                                            & 24.2\% (30.5\%)                                                                                                                                      & \small{\(\text{C}_{57}\text{H}_{91}\text{N}_{19}\text{O}_{16}\), as type I collagen }                                                                                                                \\
      \cline{3-7}
                                                          &                                         & Elastin  & 33.7\%        & 24.5\%                                                            & 19.4\% (24.5\%)                                                                                                                                      & \small{\(\text{C}_{35.94}\text{H}_{57.88}\text{N}_{11.40}\text{O}_{11.40}\), as a mixture of amino acid residues (29.8\% Glycine, 21.1\% Alanine, 13.1\% Valine, 12.7\% Proline, and 6.5\% Leucine)} \\
      \hline
      \multicolumn{3}{l}{Water}                           & 70\%                                    & 80\%     & 75\%          & \small{\(\text{H}_{2}\text{O}\)}, fraction values over total mass                                                                                                                                                                                                                                                                                                                                                               \\
      \bottomrule
    \end{tabular}
  \end{center}}
\end{table}

\begin{table}[h]
  \caption{\label{table:composition}Material compositions and densities of the key components for plaque modeling.}
  \vspace{-5truemm}
  \begin{center}
    \begin{adjustbox}{max width=\textwidth}
      \begin{tabular}{lccccccccccccc}
        \toprule
        {Material}     & Density (\(g/cm^3\)) & H      & C      & N      & O      & Na     & P      & S      & Cl     & K      & Ca     & Fe     & I      \\
        \midrule
        Vessel wall     & 1.074                & 0.1040 & 0.1818 & 0.0368 & 0.6739 & -      & 0.0011 & -      & -      & -      & 0.0024 & -      & -      \\
        Fibrous tissue  & 1.113                & 0.1089 & 0.1378 & 0.0178 & 0.7355 & -      & -      & -      & -      & -      & -      & -      & -      \\
        Necrotic core   & 1.055                & 0.1083 & 0.1737 & 0.0217 & 0.6963 & -      & -      & -      & -      & -      & -      & -      & -      \\
        Blood w. iodine & 1.060                & 0.1010 & 0.1089 & 0.0327 & 0.7376 & 0.0010 & 0.0010 & 0.0020 & 0.0030 & 0.0020 & -      & 0.0010 & 0.0100 \\
        Peri-coronary fat         & 0.950                & 0.1140 & 0.5980 & 0.0070 & 0.2780 & 0.0010 & -      & 0.0010 & 0.0010 & -      & -      & -      & -      \\
        \bottomrule
      \end{tabular}
    \end{adjustbox}

  \end{center}
\end{table}

\subsection{Simulation Studies}
\subsubsection{Deep-silicon PCCT}
Silicon is a promising sensor material for PCCT. It has a mature manufacturing process and can be made with high purity. Compared to CdTe/CZT, silicon has a higher charge carrier mobility so that it can handle higher x-ray fluxes; and it does not suffer from polarization and K-escape. On the other hand, silicon has its own limitations. Because silicon is a low-Z material, the absorption efficiency of X-rays is low and the sensors must be made very thick -- for example, a few centimeters~\cite{persson2014energy}. For this reason, GE HealthCare's deep-silicon detector is composed of silicon sheets, which are facing the X-ray source edge-on. The X-rays enter through the side of the silicon strips and travel through a long attenuation path. To cope with a high X-ray flux, the attenuation path is further divided into three segments that are read out individually. To reduce the cross-talk by X-ray scattering, there are tungsten foils between the silicon sensors.

\subsubsection{Simulation Models}

Realistic simulation is realized with CatSim~\cite{de2007catsim}, which models many detailed physics aspects of real CT scanners. For the conventional EID-CT simulation, we follow the geometry of the GE LightSpeed VCT system. For PCCT simulation, we use the parameters of the GE deep-silicon PCCT prototype. The details are summarized in Table~\ref{table:Geometry}. Note that the number of detector cells \(1500\times40\) is smaller than that used in the real PCCT scanner to save computer memory and computing time. Filtered back-projection (FBP) reconstruction was used as a common benchmark. Additionally, we used 0.35 second as a nominal example, but note that rotation speeds of up to 0.23 second per rotation are achievable in real-world CT scanners.

\begin{table}[h]
  \caption{\label{table:Geometry}CatSim simulation parameters for EID-CT and PCCT respectively.}
  \small
  \begin{center}

    \begin{tabular*}{0.9\textwidth}{l@{\extracolsep{\fill}}ll}
      \toprule
      & EID-CT & PCCT                                                           \\
      \midrule
      Source-to-isocenter distance & 541~\mm & 626~\mm\\
      Source-to-detector distance & 949~\mm & 1098~\mm\\
      Detector type & EID & PCD\\
      Total \# of detector columns & 888 & 1500\\
      Total \# of detector rows & 64 & 40\\

      Bowtie & Large & Large \\
      Focal spot size (width, length) & (0.9~\mm, 0.7~\mm) & (0.5~\mm, 0.5~\mm) \\
      Tube voltage & 100kVp & 100kVp \\
      Tube current & 1000mA & 1000mA \\

      Views per rotation & 984 & 1968\\
      Rotation period & \(+\infty\) & \(+\infty\)/0.35/0.70/2.00/2.80 sec \\
      Reconstruction algorithm & 3D FBP & 3D FBP\\
      \bottomrule
    \end{tabular*}

  \end{center}
\end{table}

To correctly reflect the attenuation and beam hardening effect in real patients, we embed our plaque model in a large adipose cylinder of 24~\cm diameter (mimicking the attenuation of an adult human body). The plaque model is offset 4.24~\cm from the center of the adipose. To minimize compute memory requirements, the adipose background phantom is represented with large voxels of 0.75~\mm in size. Additionally, to reduce aliasing artifacts induced by the low resolution phantom, a large Gaussian kernel is applied on the adipose phantom so it has very smooth boundaries.

To simulate cardiac motion, a dynamic phantom is made by laterally shifting the plaque model within the adipose in a sinusoidal pattern at a rate of 60 bpm. In this simulation study, counting mode (photon-count data are summed over all energy bins) is selected to study resolution, noise, and motion artifacts, while the spectral mode (including material decomposition) is used to characterize tissues using multiple energy bins.

\subsection{Plaque Characterization}
\subsubsection{Image Resolution}\label{sec:resolution}

To illustrate the impact of image resolution on plaque characterization, we simulate the digital phantom under noise-free conditions with EID-CT and PCCT respectively. The noise-free image demonstrates the resolution limit that a practical image can approach but we often need to sacrifice some resolution to gain better signal quality in presence of noise. \mz{The residual motion artifacts could also reduce the final resolution.} To find out the resolution margin that could be \mz{reserved} for noise compensation and residual motion artifacts, we incrementally blur the PCCT images with Gaussian kernels to investigate the critical resolution point that is still acceptable for plaque imaging. Correct measurement of the fibrous cap thickness and the lumen area within 20\% error serves as the criteria for acceptable images. All images are reconstructed with a standard kernel and isotropic voxels of size \(0.1~mm\times 0.1~mm\times 0.1~mm\), resulting a \(90\times 90\times 113\) volume. Note that although this is not a normal reconstruction voxel size, it is used to study the system resolution in an extreme case. For better visualization and further analysis, 2D cross-section images are then linearly interpolated by a factor 2.84 (\(0.035^2~mm^2\) pixels). The lumen is segmented with a threshold of 250 HU which is half the normal value of contrast enhanced blood region within lumen. The necrotic core region is identified by selecting the pixels with values between 20 HU and 60 HU corresponding to 0.5 and 1.5 times the normal value (40HU). The area of lumen is calculated by counting the number of pixels from the segmented lumen. The thickness of the fibrous cap is measured as the minimum distance between the lumen segment and the necrotic core region.

\subsubsection{Image Noise}\label{sec:noise}
Subsection~\ref{sec:resolution} illustrates the impact of image resolution in the ideal scenario, but in practice a clinical image is achieved in a balance between noise and resolution. Various combinations of reconstruction kernels and voxel size/slice thickness provide such options for optimized image quality. In this study, kernels of soft, standard, detail and bone types and voxel size of 0.2~\mm, 0.25~\mm, 0.3~\mm, and 0.4~\mm~are investigated to find the best combination. 
Since CatSim currently only supports analytic reconstruction methods while iterative reconstruction and deep learning methods are widely used in practice for PCCT, here we increase the simulated dose up by 5.7x to mimic the use of advanced denoising techniques (corresponding to 82\% radiation dose reduction) such as iterative reconstruction~\cite{si2022coronary,silva2010innovations} or deep learning techniques. Realistically, denoising techniques are said to be equivalent to about 1.7 to 5 times dose reduction (or 40\% to 80\% radiation dose reduction) for iterative reconstruction techniques, and additional 2 to 4 folds reduction to be expected (or another 50\% to 75\% radiation dose reduction) if advanced deep-learning-based methods are applied. For example, \cite{mathieu2014radiation} suggests 40\% and 80\% radiation dose reduction can be achieved from adaptive statistical iterative reconstruction (ASIR) and model-based image reconstruction (MBIR) respectively relative to FBP reconstruction, while maintaining similar image noise levels and contrast-to-noise ratios for lung cancer screening. Similarly, \cite{chang2013assessment} with MBIR half-dose liver CT (46.1\% dose reduction) demonstrates superior CNRs and less image noise than the reference full-dose CT. The associated dose reduction factor with MBIR was reported as 74\% in~\cite{herin2015use} for reduced-dose body CT compared with the standard dose CT with FBP and 79\% in~\cite{katsura2012model} for low-dose chest CT compared with the reference dose CT with ASIR. \cite{samei2015assessment} suggests a dose reduction range between 47\% and 89\% for MBIR relative to FBP depending on tasks. Both a review paper~\cite{kataria2021image} and a research study~\cite{ellmann2018advanced} pointed out advanced modeled iterative reconstruction (ADMIRE) Strength 3 allows a 30\% patient radiation dose reduction relative to FBP while~\cite{solomon2015diagnostic} shows a radiation dose reduction between 4\% and 80\% with ADMIRE while preserving low-contrast detectability. Similar dose reductions were also observed with iterative reconstruction techniques developed by other vendors, such as sinogram affirmed iterative reconstruction (SAFIRE)~\cite{kalra2012radiation,solomon2017effect} and adaptive iterative dose reduction (AIDR)~\cite{gervaise2012ct,yamada2012dose}. More recently, the emerging deep learning-based image reconstruction techniques, such as the TrueFidelity from GE HealthCare and Advanced Intelligent Clear-IQ Engine (AiCE) from Canon Medical Systems, enable a greater radiation dose reduction potential compared to the classic iterative reconstruction methods. For instance, \cite{noda2021low} shows a radiation dose reduction over 75\% for whole-body CT with TrueFidelity compared to standard-dose CT with ASIR-Veo and iDose (a hybrid iterative reconstruction method by Philips Medical Systems) while maintaining similar lesion detection rates and image quality measures. Futhermore, \cite{brady2021improving} demonstrates a radiation dose reduction factor of 52\% with AiCE relative to the AIDR-3D method.

To achieve dose neutrality, here we claim a conservative 5.7 times dose reduction with advanced deep learning techniques such as MBIR combined with deep learning prior or dual-domain denoising. Additionally, we also evaluate a dose-optimized bowtie design for ultrahigh resolution volume of interest imaging~\cite{robinson2016experimental}, which will be further justified in Section~\ref{sec:dose}. With this bowtie technique accounting for 2x dose reduction (or 50\% dose reduction), we can safely claim a more conservative 4x dose reduction (or 75\% dose reduction) with deep learning techniques for a total of 8x dose reduction.
 
Besides the qualitative visual evaluation, we also use the detectability of the necrotic core as our quantitative criterion to single out the best reconstruction. Finally, the exemplary reconstructions under various noise strength scenarios are demonstrated to showcase the expected dose neutral image quality with a 8x dose reduction factor brought by a advanced deep learning denoiser and the ROI-oriented bowtie.

For quantitative measurement, we repeat the simulation for 10 realizations with random noise, and use the contrast to noise ratio (CNR) as our metric, which is calculated as the ratio between the mean difference of background against foreground regions of interest (ROIs) and the noise captured as the standard deviation of means of the ROIs. The foreground ROI is specified within the necrotic core while the background ROI locates in the surrounding fibrous tissue.

\begin{equation}
  CNR = \frac{\mu_{bg} - \mu_{fg}}{\sqrt{\sigma_{bg}^2 + \sigma_{fg}^2}},
\end{equation}
where \(\mu_{(\cdot)}\) and \(\sigma_{(\cdot)}\) denote the mean and standard deviation values over the 10 realizations of the mean of the background ROI (\(bg\)) or foreground ROI (\(fg\)). Note that the noise estimation does not correspond to the actual noise level in a single image realization after the average operation over multiple pixels, and its value is also affected by the size of the ROI. For fair comparison, all reconstructions from different voxel sizes are linearly interpolated to the same voxels to maintain a fixed ROI size. Also, it is challenging to delineate the ROIs on a reconstruction with coarse voxels due to the small size of the plaque. In this study, the interpolation voxel size is \(0.035~mm \times 0.035~mm \times 0.035~mm\), leading to an axial image size of \(256\times256\).

\subsubsection{Motion Artifacts}
A dynamic phantom with 1,100 phases in one cycle is used in our dynamic simulation study. The in-plane motion follows a temporally sinusoidal pattern of 2~\mm~in amplitude and 60bpm~\cite{zsarnoczay2023ultra}. To investigate the tolerance to residual motion artifacts in plaque imaging, the motion amplitude is reduced to 10\%, 20\%, 40\%, and 60\% the original respectively. Since the rotation period of the scanner is set to 0.35 seconds per turn, the amplitudes correspond to a movement range of [0, 0.2~\mm], [0, 0.4~\mm], [0, 0.8~\mm], and [0, 1.2~\mm] respectively.
The images are then reconstructed with 0.2~\mm~voxels under the noise-free condition.
To quantitatively study the impact of residual motion artifacts, the lumen and necrotic core regions are segmented with the same thresholding method described in Subsection~\ref{sec:resolution}. The area of the segmented lumen is also calculated under various levels of residual motion artifacts.

\subsubsection{Spectral Characterization}
\mz{Virtual monochromatic images and material maps are used to demonstrate the spectral imaging capability of PCCT. We first perform material decomposition in the projection domain and FBP to reconstruct basis material images, then use Formula~\ref{eq:CTE} to generate virtual monochromatic images. PE and PVC are selected as the basis materials. The basis material images and virtual monochromatic images are showing in Fig.~\ref{fig:SpectralCNR}(a)-(e).
\begin{equation}
    CT(E) = 1000\frac{L_1\mu_{1,E}+L_2\mu_{2,E} - \mu_{water,E}}{\mu_{water,E} - \mu_{air,E}},\label{eq:CTE}
\end{equation}
where \(E\) is the the energy of the monochromatic image, \(L_1\) and \(L_2\) are the basis material image values, \(\mu_{(\cdot,E)}\) denotes the linear attenuation coefficient at energy \(E\). We calculate the contrast between the necrotic core and the fibrous tissue from the noise-free virtual monochromatic images, and estimate the background noise from the noisy images, as the function of energy, then find the CNR as showing in Fig.~\ref{fig:SpectralCNR}(f). The optimal energy is 52 keV.
\begin{figure}
  \centering
  \includegraphics[width=0.8\linewidth]{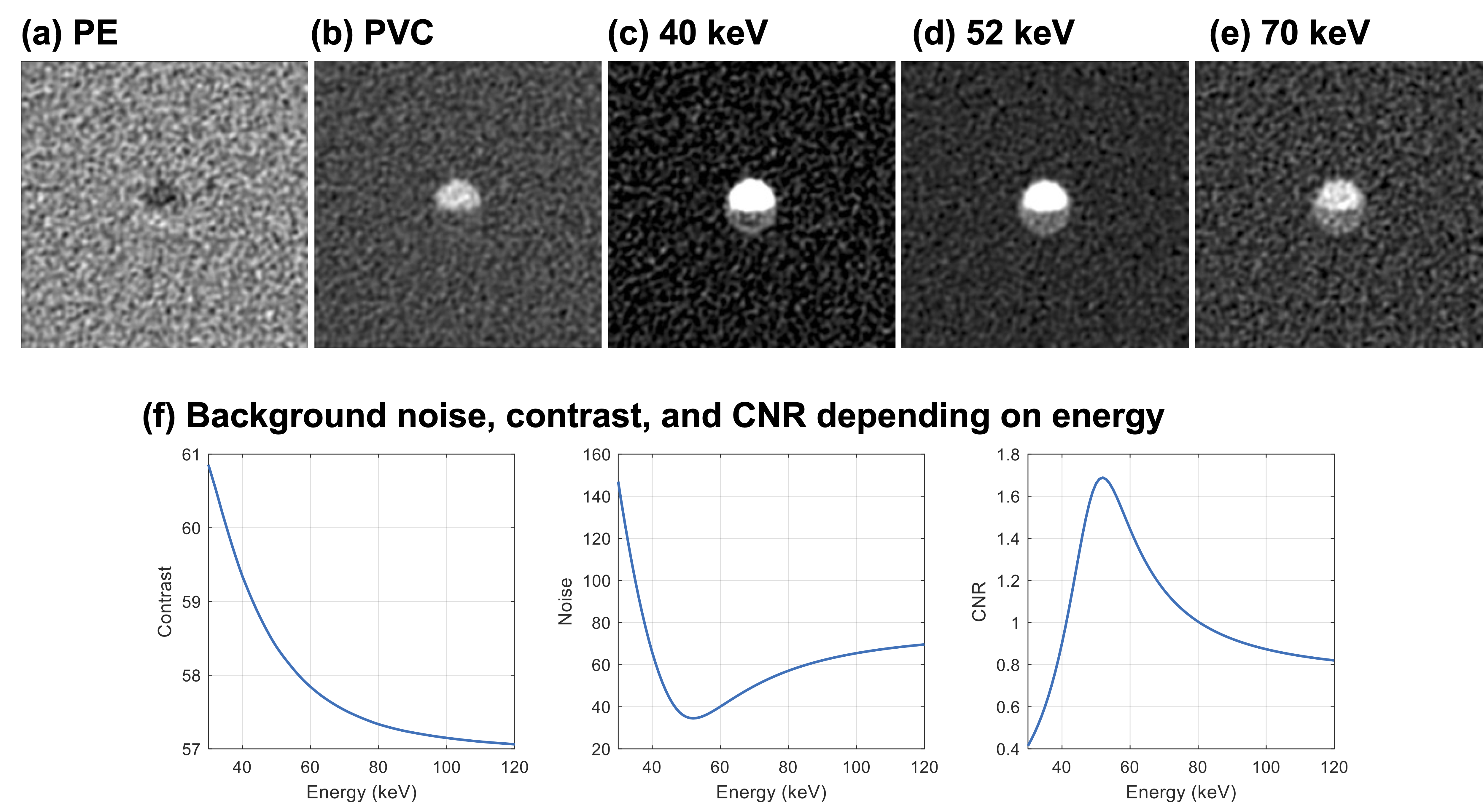}
  \caption{\mz{Basis material images: (a) and (b); Virtual monochromatic images: (c) – (e); Background noise, contrast between the necrotic core and the fibrous tissue, the Contrast-to-Noise-Ratio (CNR) as the function of the monochromatic energy.}}\label{fig:SpectralCNR}
\end{figure}

A four-material image domain material decomposition method was implemented in this work to differentiate tissue types of PCCT plaque images. We use the analytical multi-material decomposition (MMD) method~\cite{mendoncca2013flexible} to decompose the PE and PVC basis images into four materials: i.e., iodinated blood, necrotic core, fibrous tissue, and peri-coronary fat (detailed composition of these material are described in Sec.~\ref{sec:composition}). A standard 2D k-means clustering algorithm was used to reduce noise in MMD results. All material decompositions were performed on simulated plaque images at 0.2~\mm~isotropic voxel size under three different noise levels (25\% standard noise, 50\% standard noise, and standard noise). The ability to distinguish different plaque tissue types (especially the necrotic core) is qualitatively evaluated.}

\subsubsection{Radiation Dose}\label{sec:dose}
Usually, more photons are required for higher resolution imaging to combat the increased image noise, which scales quadratically with spatial resolution~\cite{hirano2004relationship,brooks1976statistical}. On the other hand, radiation dose directly relates to the number of penetrating photons through human body. While the radiation dose is harmful, we can adopt a series of techniques to reduce the radiation dose of our high-resolution PCCT scan. For example, we can substantially cut the radiation dose using contemporary deep learning techniques for denoising and deblurring, which have been widely reported in the literature~\cite{silva2010innovations,noda2021low,parakh2021sinogram,niu2022noise} and also justified in Section~\ref{sec:noise}. Here we present another major conceptual idea to be considered in practice for further dose optimization in plaque imaging applications. We propose to utilize an ROI bowtie to confine a high X-ray flux only to the cardiac region while suppressing the X-ray flux outside the ROI. With this technique, we can maximize the tube current of a small-focus X-ray source but still keep the radiation dose at a low level by optimizing the radiation dose deposition. Here we focus on the simulation demonstration of dose reduction with an ROI bowtie. We follow the idea presented in~\cite{robinson2016experimental} but instead of showing a dynamic bowtie we simulate an equivalent static bowtie assuming an ROI centered at the iso-center for simplicity. The diameter of the ROI is set to about 10~\cm. The thin region of the ROI bowtie reaches zero in thickness while the thick region is 3.9~\cm to block x-rays. The shape of the ROI bowtie is illustrated in Fig.~\ref{fig:HumanDose}(b) assuming at a position of 26.3~\cm from the iso-center. To demonstrate the dose reduction effect of the ROI bowtie, we compared the results from PCCT scans with and without using the technique; i.e., PCCT scans with a conventional large bowtie versus PCCT scans with a conventional large bowtie plus an ROI bowtie. Since the radiation dose simulation is irrelevant to X-ray detection, we used an EID for detection with electronic noise turned off and 0.78~\mm reconstruction voxels for simplicity, and followed the other acquisition parameters illustrated in Table.~\ref{table:Geometry}; e.g., 1000~\(mA\) tube current, a rotation time of 0.35 seconds and 1968 projection views. Without loss of generality, two dimensional (2D) studies were performed with a human thorax slice and a 32~\cm CT dose index (CTDI) phantom for qualitative comparison and quantitative evaluation respectively. The absorbed dose reconstruction was simulated with the open source X-ray-based Cancer Imaging Toolkit (XCIST)~\cite{wu2022xcist}. The 2D human thorax phantom was included in XCIST as the adult male 50 percentile chest phantom slab, while the CTDI phantom composes of a PMMA body of 320~\mm~in diameter and four water inserts of 13~\mm~in diameter. The beam hardening effect was corrected with a fifth order polynomial function, and a series of calibration thicknesses in a 10~\mm~step size up to a 380~\mm~maximum thickness.

\section{Results}
\subsection{Resolution Consideration}

\begin{figure}
  \centering
  \includegraphics[width=0.9\linewidth]{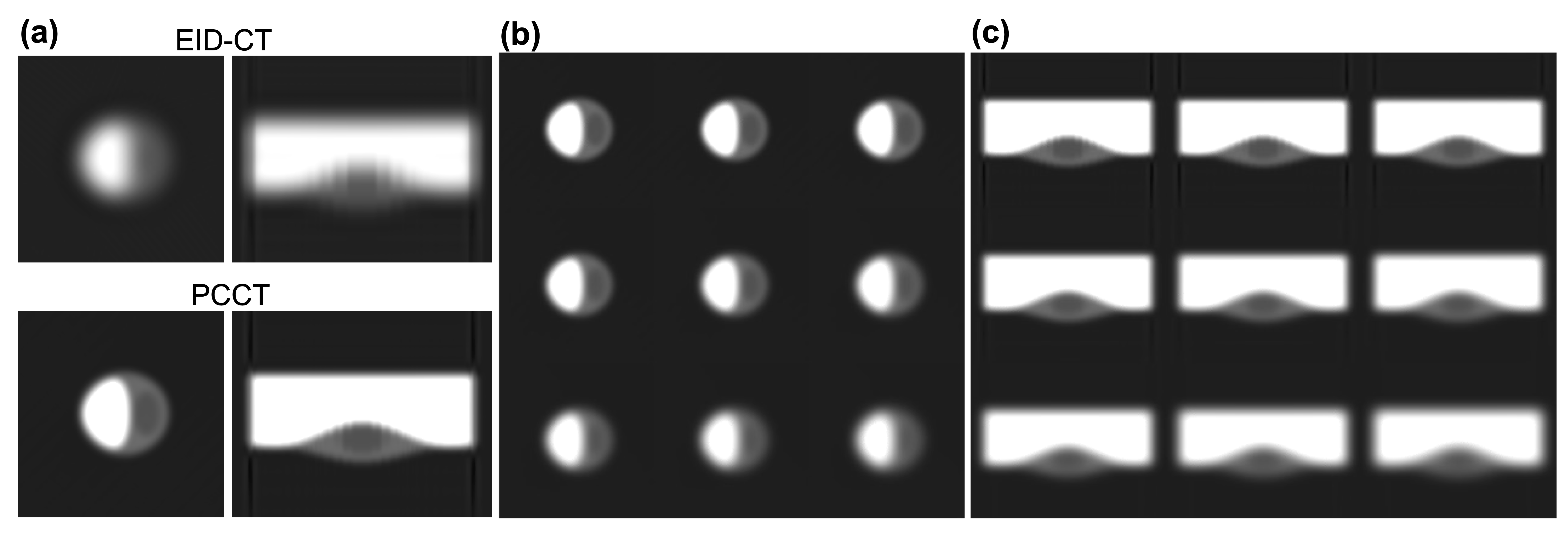}
  \caption{Noise-free plaque imaging in different resolutions. (a) EID-CT (top row) versus PCCT (bottom row); (b) and (c) PCCT imaging with resolution degradation at different levels with various Gaussian blurring kernels (from 0 to 10 in standard deviation). All images are in the same display window (C150 HU/W600 HU).}\label{fig:SECTvsPCCT}
\end{figure}

The central axial and sagittal cross-sections of the PCCT reconstruction are displayed against the EID-CT counterparts in Fig.~\ref{fig:SECTvsPCCT}(a) with window and level being 150 HU and 600 HU respectively. The PCCT images are further degraded with various Gaussian blurring kernels from 0 to 10 in standard deviation \mz{(in unit of 0.035~\mm-pixel)} as shown in Fig.~\ref{fig:SECTvsPCCT}(b) and (c).
Fig.~\ref{fig:SECTvsPCCT}(a) illustrates the case of plaque imaging with an EID-CT scanner where the necrotic core is hardly discernible due to insufficient spatial resolution limited by a large detector pixel size and a matching focus spot. Although a ``napkin-ring'' sign is possibly visible~\cite{otsuka2013napkin}, the boundary of the core cannot be resolved which limits a reliable measurement of the cap lid thickness for risk stratification of the plaque. On the other hand, in the PCCT images with a much finer resolution, the necrotic core boundary is well delineated with a good visibility of the cap lid. A somehow staircase-like boundary of the lumen in the sagittal view is caused by the significant difference between the in-plane and through-plane resolution due to a greater dimension in the row direction of a detector cell than that in the column direction. A slight reduction in sharpness could help remove the issue as suggested in Fig.~\ref{fig:SECTvsPCCT}(c). Note that the PCCT images in Fig.~\ref{fig:SECTvsPCCT}(a) were reconstructed at a small voxel size without considering quantum noise and crosstalk blurring between pixels, and they reflect the best possible result that could be approached in practice. This ideal case resolution is often compromised in practice to compensate for other image quality aspects such as noise level. To investigate the allowable range of this margin, we degraded the resolution of PCCT images with a series of Gaussian blurring kernels characterized by 9 uniformly distributed standard deviation values from 0 to 10. The resultant images in Fig.~\ref{fig:SECTvsPCCT}(b) and (c) present the gradually decreasing visibility of the necrotic core. Visually, the 4th and 5th sets of the images still show relatively clear necrotic core boundaries, which correspond to a kernel with a standard deviation between 3.75 and 5. 
Hence, the maximum acceptable resolution margin for clear visibility of the necrotic core boundary can be measured as the full width at half maximum (FWHM) of the Gaussian kernels in that range, which is around 0.31~\mm~to 0.41~\mm.

\begin{figure}
  \centering
  \includegraphics[width=0.75\linewidth]{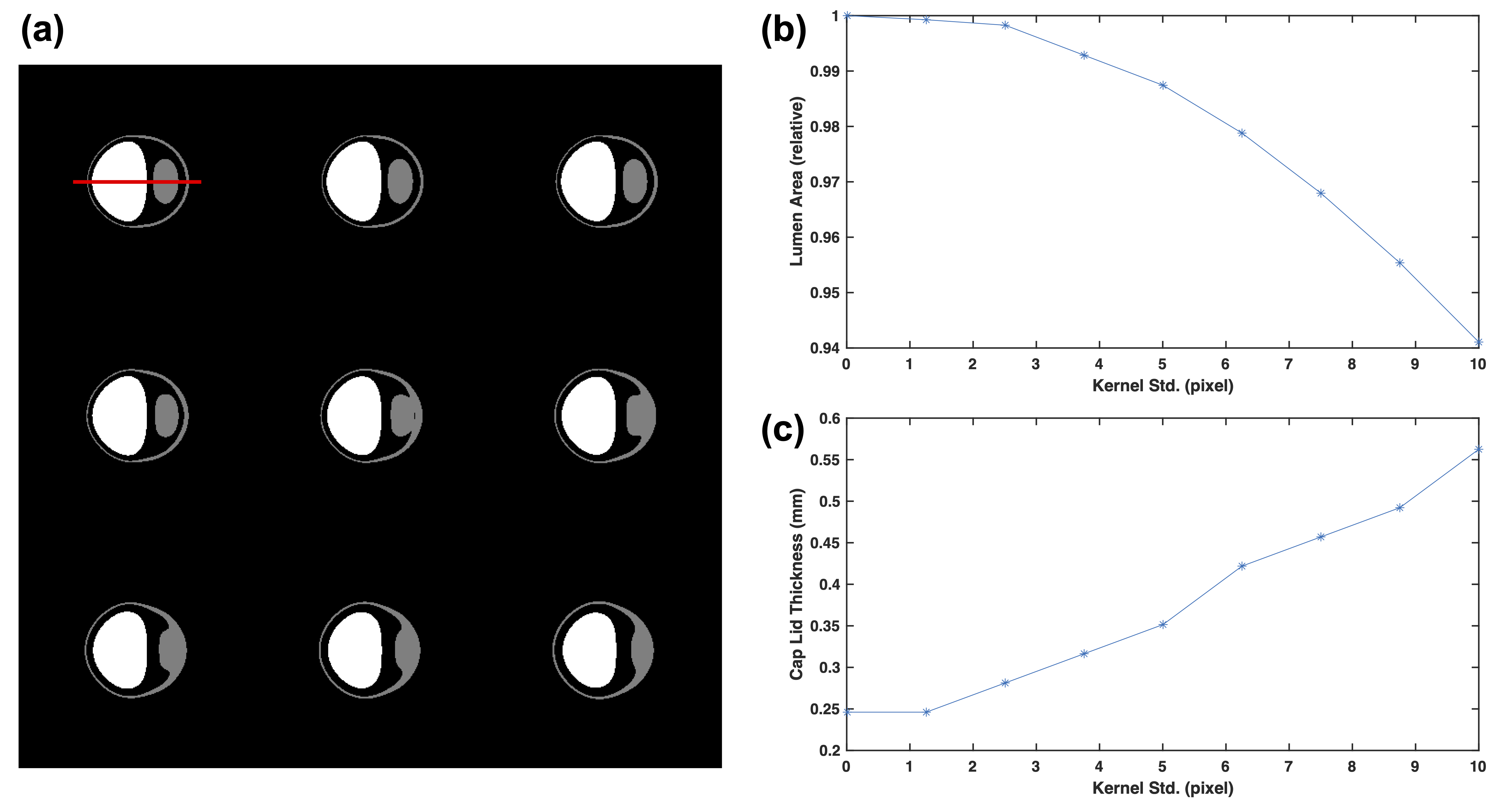}
  \caption{Segmentation maps for Fig.~\ref{fig:SECTvsPCCT}(b) and corresponding measurements of lumen area and cap lid thickness. (a) The thresholding-based segmentation for lumen (white) and necrotic core (gray); (b) the normalized area of the lumen segmentation as a function of the blurring kernel size, and (c) the cap lid thickness measurement gradually deviates from the correct value as image resolution reduces.}\label{fig:SegMap}
\end{figure}

The lumen (white) in Fig.~\ref{fig:SECTvsPCCT}(b) is segmented in Fig.~\ref{fig:SegMap}(a) accompanied with the necrotic core colored in gray. The curve of the measured lumen area against the blurring kernel size is normalized with the original PCCT result and displayed in Fig.~\ref{fig:SegMap}(b). The curve of the cap lid thickness against the blurring kernel size is plotted in Fig.~\ref{fig:SegMap}(c). As shown in Fig.~\ref{fig:SegMap}(a), the lumens are well segmented for all the blurring kernels with only a subtle shrinkage of the shape, which is also reflected as a small change of the measured lumen area in Fig.~\ref{fig:SegMap}(b). Interestingly, unlike the blooming effect that makes the dense object appear bigger, given proper prior knowledge of the CT values, decreasing resolution leads to smaller lumen segmentation results. Compared to the lumen, the segmentation of the necrotic core is more strongly influenced by the resolution degradation. The necrotic cores were well identified and separated from the vessel media for the first four kernels while the segmentation was significantly deformed as the blurriness further increases, as shown in Fig.~\ref{fig:SegMap}(a). Similarly, the cap lid thickness measurement (measured as the minimum distance between the lumen and the necrotic core along the red horizontal line illustrated in Fig.~\ref{fig:SegMap}(a)) is more sensitive to the resolution change, as demonstrated in Fig.~\ref{fig:SegMap}(c). When taking 20\% error as the accepting threshold, the kernel size should be less than 3.75 in standard deviation, which corresponds to a 0.31~\mm~resolution margin threshold in terms of FWHM.

Combining the above two considerations (\(<\)0.31~\mm) and the system resolution, the final resolution is recommended to be better than 0.43~\mm~to allow accurate plaque imaging for measurement of both lumen area and cap lid thickness.

\subsection{Noise Influence}

\begin{figure}
  \centering
  \includegraphics[width=0.9\linewidth]{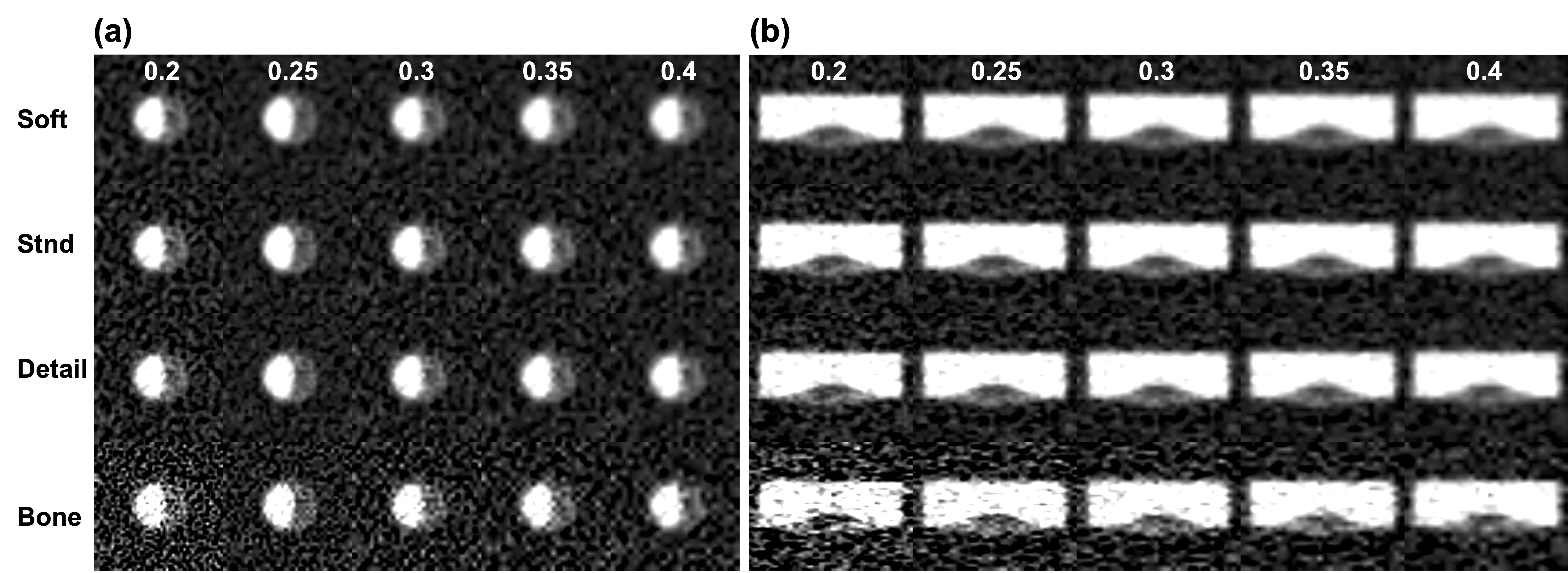}
  \caption{Visualization of the plaque image reconstructions with various combinations of kernels and voxel sizes in the (a) axial and (b) sagittal views. The reconstruction kernels from the top to bottom row are soft kernel (Soft), standard kernel (Stnd), detail kernel (Detail), and bone kernel (Bone) while different reconstruction voxel sizes from 0.2~\mm~to 0.4~\mm~denoted on the top of the images. The window level and width are the same for all the images (C150 HU/W600 HU).}\label{fig:NoiseAndVoxel}
\end{figure}

\begin{figure}
  \centering
  \includegraphics[width=0.9\linewidth]{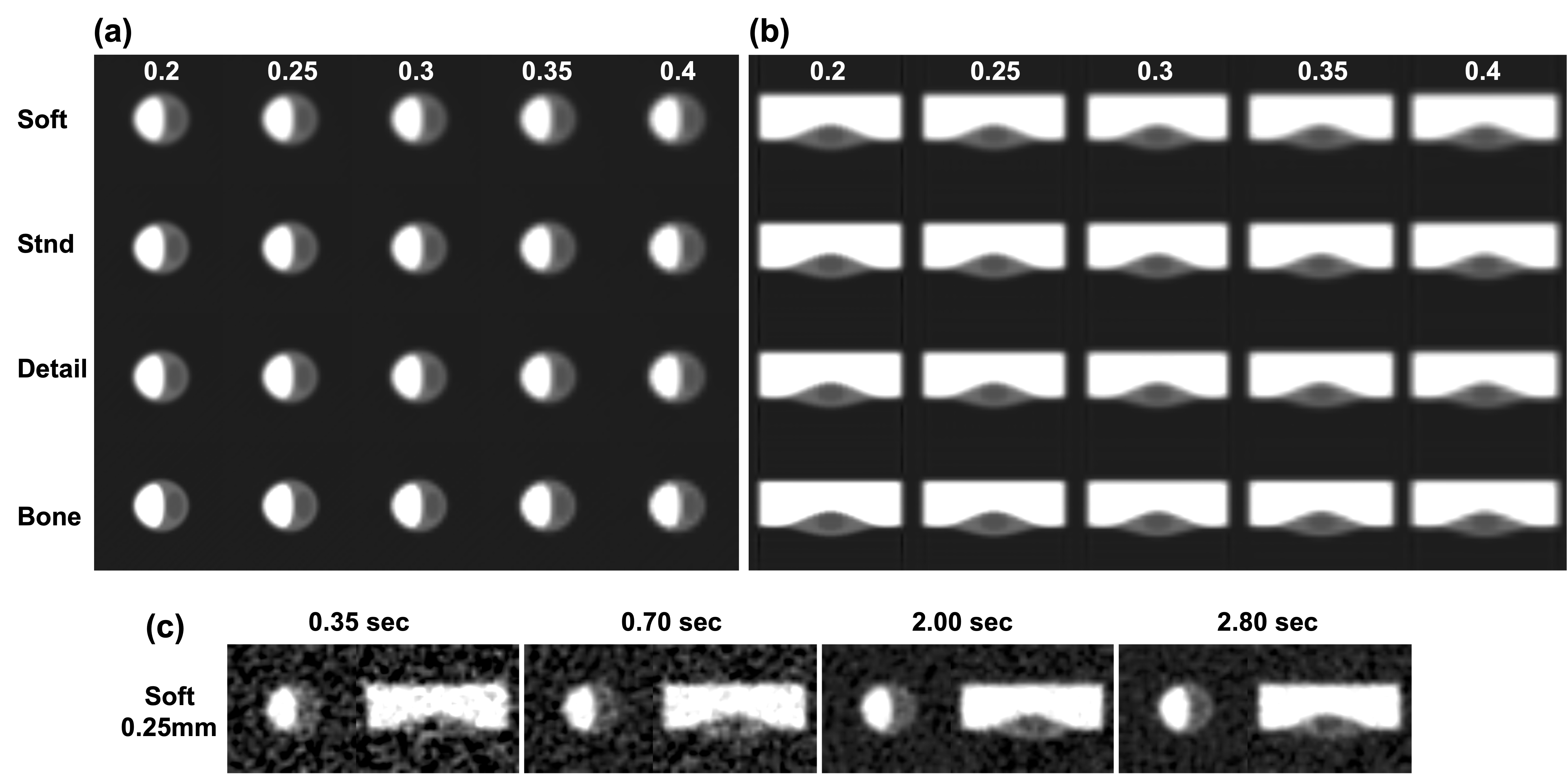}
  \caption{Noise-free version of Fig.~\ref{fig:NoiseAndVoxel} (a) and (b), with (c) showing reconstruction results with the soft kernel and 0.25~\mm~voxels under various noise levels (fixed tube current with four different exposure time settings). The window level and width are the same for all the images (C150 HU/W600 HU).}\label{fig:NoiseAndVoxel_noisefree}
\end{figure}

Exemplary reconstruction results of one realization are illustrated in Fig.~\ref{fig:NoiseAndVoxel} with four standard kernels and five reconstruction voxel sizes respectively. The noise-free version is displayed in Fig.~\ref{fig:NoiseAndVoxel_noisefree}. It is easy to see the huge impact of noise on the image features of the plaque by comparing Figs.~\ref{fig:NoiseAndVoxel} and~\ref{fig:NoiseAndVoxel_noisefree}. While the plaque necrotic cores are visible in all noise-free reconstructions, some of them are buried in noise textures in the noisy reconstructions. The trend is also evident that sharper kernels (detail and bone kernels) lead to stronger noises, which is expected and shown in Fig.~\ref{fig:NoiseAndVoxel}. 
Additionally, Fig.~\ref{fig:NoiseAndVoxel_noisefree} shows that the structural details begin to lose when the voxel becomes larger than 0.35~\mm, suggested by the blockiness in the presentation e.g., the pixelated blotchy appearances of vessel boundaries in the 4th and 5th columns. In the presence of noise, the soft kernel provides the best visual appearance with the lowest noise level as compared to the other kernels while the 0.25~\mm-voxel reconstructions demonstrate the most appealing results with a decent preservation of feature shapes at a low noise level. Note the small dark holes near the plaque center in Fig.~\ref{fig:NoiseAndVoxel}(a) do not necessarily represent the necrotic cores as they are caused by noise which do not reflect the correct size or value suggested by the noise free versions in Fig.~\ref{fig:NoiseAndVoxel_noisefree}(a). Overall, the soft kernel and standard kernel reconstructions of 0.25~\mm~voxels achieve the best balance between noise and detail, providing visually the most desirable results.

\begin{figure}
  \centering
  \includegraphics[width=0.6\linewidth]{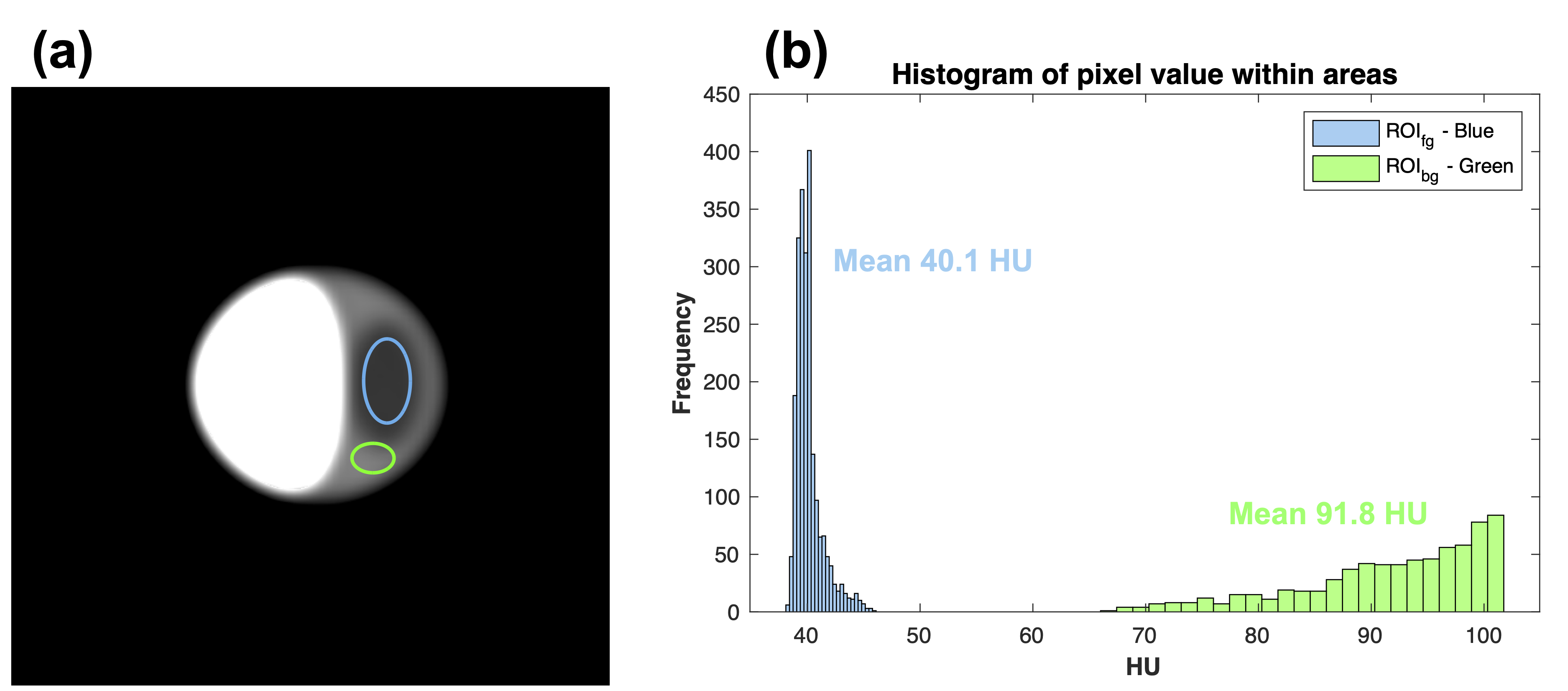}
  \caption{Illustration of the ROI locations and the corresponding histogram of pixel values within the ROIs. (a) The foreground ROI (blue) and background ROI (red) delineated on the noise-free reconstruction of \(0.1^3mm^3\) voxels with the Stnd kernel (C100 HU/W200 HU), and (b) the histograms of pixel values within the two ROIs.}\label{fig:ROI_illustration}
\end{figure}

To quantify these results, Fig.~\ref{fig:ROI_illustration} illustrates the ROI positions we used for calculation and exemplary histogram of pixel values within the ROIs for a noise-free reconstruction with the standard kernel and 0.1~\mm~voxels. The good separation in the histogram between two ROI indicates the valid positioning of the ROIs (they are fully within the necrotic core region and the fibrous tissue region respectively). The mean value of the necrotic core ROI is 40.1 HU while the fibrous tissue ROI has a mean value of 91.8 HU. The latter ROI is placed close to the necrotic core boundary to reflect the influence of resolution, leading to a slight deviation of its mean value from 100 HU. Since this reconstruction is noise-free and has the highest resolution, the mean value of the ROI could serve as the reference for reconstructions with noise and large voxels. The CNR values are measured based on the ROIs, and the statistics of the 10 noisy realizations are in Table~\ref{table:Statistics} with the corresponding CNR values, as shown in Table~\ref{table:CNR}. The smallest standard deviation values for 0.25~\mm~voxels agree with our visual observation that the 0.25~\mm-voxel reconstruction provides the lowest noise level. Furthermore, their mean values are the closest to the reference value around 40 HU and 92 HU for the necrotic core and the fibrous tissue respectively, suggesting a good balance between resolution and noise strength. Similar to our visual observations, the combination of 0.25~\mm~voxels and soft or standard kernels yield the top two CNR values, providing the best necrotic core detectability.

The reconstructions with the soft kernel and 0.25~\mm~voxels under various noise levels are displayed in Fig.~\ref{fig:NoiseAndVoxel_noisefree}(c), where four noise levels correspond to 1x, 2x, 5.7x, and 8x radiation dose strengths controlled by exposure time. The base scan with a 0.35-second scan time offers a result too noisy for reliable discrimination of the necrotic core. Similar results are observed in the 0.70-second scan with doubled dose, while the noise in the 2.0-/2.8-second results is less significant and the necrotic core structures are depicted with sufficient details for reliable detection, showcasing the expected image quality baseline for PCCT plaque imaging with advanced deep learning techniques.

\begin{table}[h]
  \caption{\label{table:Statistics}Statistics on the means of the ROI over 10 realizations under various reconstruction settings (with the standard deviation in parentheses).}
  \small
  \begin{center}

    \begin{tabular*}{0.9\textwidth}{l@{\extracolsep{\fill}}cccccc}
      \toprule
      \multicolumn{2}{c}{Recon Voxel (\mm)} & 0.20 & 0.25 & 0.30 & 0.35 & 0.40                                                          \\
      \midrule
      \multirow{4}{*}{\(ROI_{fg}\)} & soft &48.76 (11.93) & 43.47(9.73) &  49.62(10.76) &   49.30 (10.33)&   51.56(10.44)\\
      & stnd & 50.05(14.29)&  43.04(10.54)&   49.86(12.44)&   48.97(11.52)&   50.92(12.42)\\
      & detail &50.58(14.19)&   43.91(10.45)&   50.12(13.09)&   49.43(11.54)&   51.39(12.77)\\
      & bone &51.26(17.56)&   44.38(12.05)&   47.48(22.38)&   48.14(14.61)&   50.35(20.36)\\
      \midrule
      \multirow{4}{*}{\(ROI_{bg}\)} & soft& 88.38(27.37)&   86.44(16.91)&   84.07(24.28)&   83.48(21.43)&   82.12(17.30)\\
      &stnd &97.51(34.28)&   93.10(21.92)&   90.86(29.60)&   90.01(24.49)&   88.40(18.25)\\
      &detail &98.40(34.55)&   93.07(23.35)&   90.17(31.22)&   91.28(24.56)&   88.51(16.58)\\
      &bone &111.06(47.88)&   99.80(39.66)&   91.62(52.88)&  104.60(32.34)&   91.97(17.85)\\
      \bottomrule
    \end{tabular*}

  \end{center}
\end{table}

\begin{table}[h]
  \caption{\label{table:CNR}Necrotic core detectability measured as the CNR (with the highest value in bold).}
  \small
  \begin{center}

    \begin{tabular*}{0.6\textwidth}{l@{\extracolsep{\fill}}ccccc}
      \toprule
      {Recon voxel (mm)} & 0.20 & 0.25 & 0.30 & 0.35 & 0.40                                                          \\
      \midrule
      soft &1.33  & \(\bm{2.20}\) & 1.30 & 1.44 & 1.51\\
      stnd & 1.28 & 2.06 & 1.28 & 1.52 & 1.70 \\
      detail&1.28 & 1.92 & 1.18 & 1.54 & 1.77 \\
      bone &1.17 & 1.34 & 0.77 & 1.59 & 1.54\\
      \bottomrule
    \end{tabular*}

  \end{center}
\end{table}

\subsection{Motion-affected Results}

\begin{figure}
  \centering
  \includegraphics[width=0.8\linewidth]{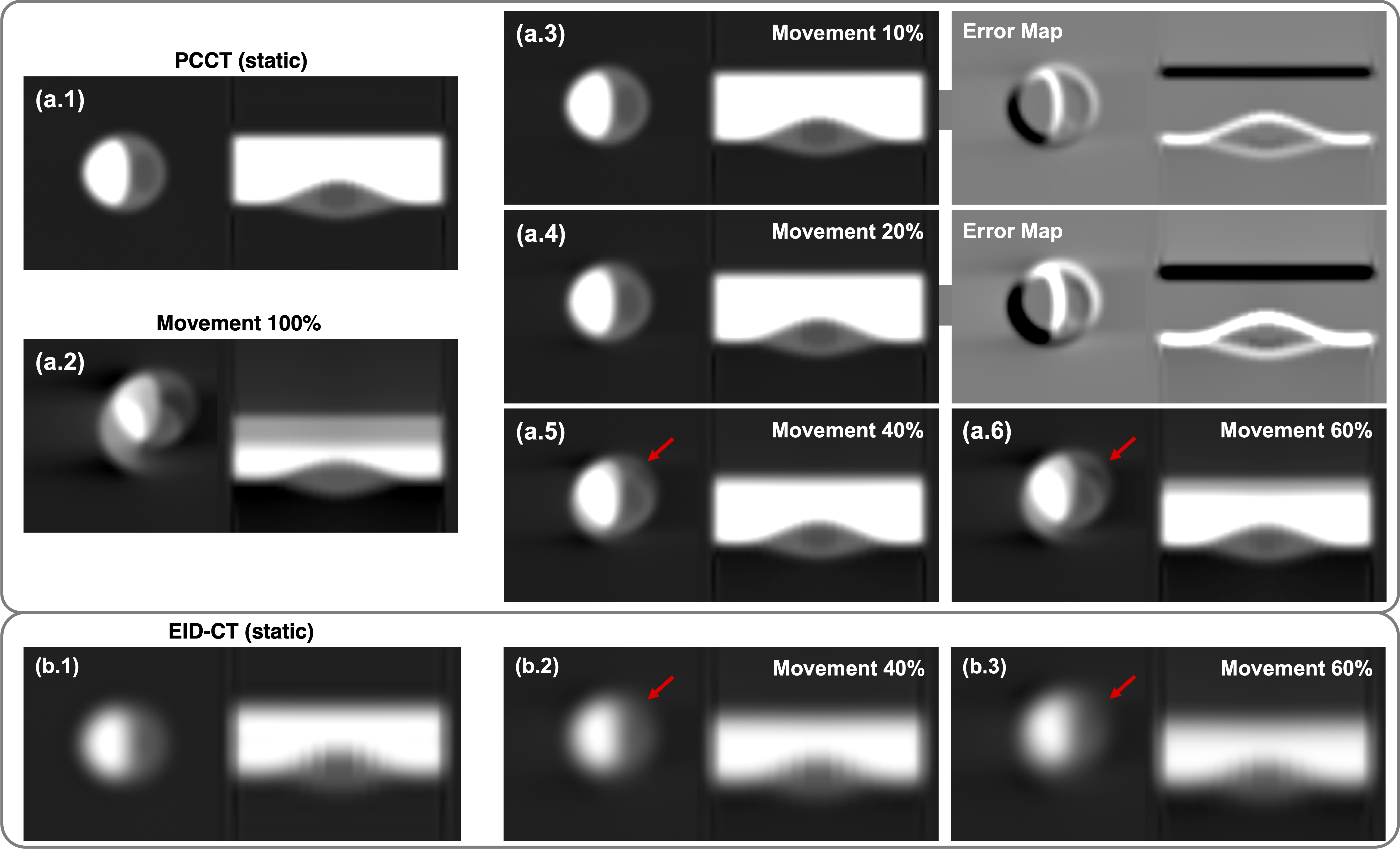}
  \caption{Impact of cardiac motion on plaque image quality. The top group demonstrates the PCCT result from (a.1) a static phantom as the reference, (a.2) a dynamic phantom with a full motion amplitude (2~\mm~in-plane movement following a 60bpm sinusoidal pattern and the rotation period of the scanner is 0.35 second), and (a.3 to a.6) reconstruction results with fractional motion amplitudes (10\%, 20\%, 40\%, and 60\%). The results in (a.3) and (a.4) demonstrate less visible motion artifacts, accompanied with corresponding error maps on the right showing positional shifting effect. The bottom group displays (b.1) the reference EID-CT result with the static phantom, as well as (b.2) and (b.3) results from the dynamic phantom with fractional motion amplitudes (40\% and 60\%). The red arrows point to the motion artifacts that are evident in the PCCT result but less noticeable in the EID-CT result due to the resolution difference. The window level and width for the reconstruction results and error maps are C150 HU/W600 HU and C0 HU/W200 HU respectively.}\label{fig:Motion}
\end{figure}

A major issue with cardiac CT is motion artifacts. The motion-affected results for PCCT are displayed against the static reference and accompanied with error maps in Fig.~\ref{fig:Motion}(a). With a strong movement pattern (100\% movement magnitude), the plaque and lumen structures are not correctly reconstructed. Then, they are significantly improved as the movement amplitudes are reduced to 60\%, where the plaque necrotic core begins to appear in the sagittal view though major artifacts are presented in the axial view. As the movement gets further corrected, the motion artifacts become less prominent. For example, with 40\% movement the artifacts are still visible but with movement magnitude below 20\% the artifacts become insignificant in reconstruction images due to its limited complication (20\% magnitude corresponds to a 0.4~\mm~movement range at maximum, which is around 2-pixel width on the detector plane) although the motion effects can be still observed in the error map as translated structures. When comparing between the PCCT and EID-CT reconstructions of the dynamic phantom with 40\% movement, the EID-CT images are less sensitive to the motion and their appearances are almost the same as those with a static phantom shown in Fig.~\ref{fig:Motion}(b.1). Similarly, the motion induced structural artifacts are more conspicuous in PCCT than those in EID-CT for dynamic phantom with 60\% movement as revealed in Figs.~\ref{fig:Motion}(a.6) and (b.3). This difference in tolerance suggests that higher resolution PCCT has stricter requirements on correction of residual cardiac motion artifacts.

\begin{figure}
  \centering
  \includegraphics[width=0.8\linewidth]{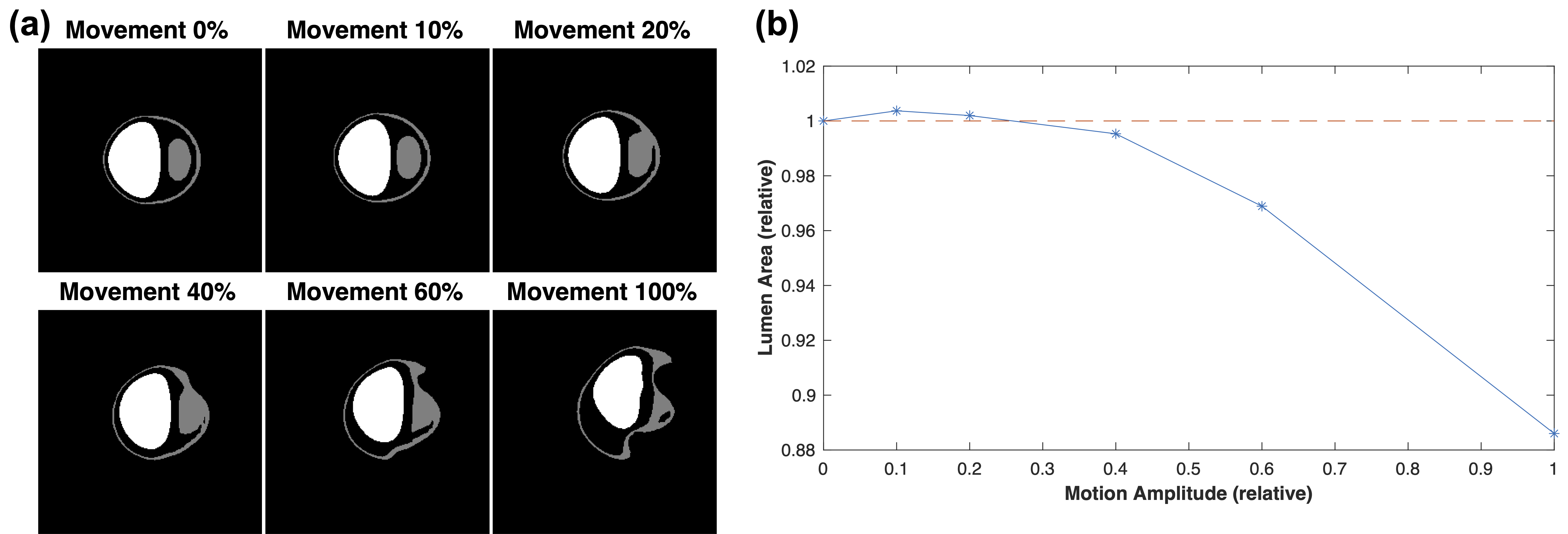}
  \caption{Segmentation maps for PCCT images with cardiac motions in Fig.~\ref{fig:Motion} and corresponding measurements of lumen area. (a) The thresholding-based segmentation for lumen (white) and necrotic core (gray); (b) the normalized area of the lumen segmentation as a function of the residual motion amplitude.}\label{fig:MotionAnalysis}
\end{figure}

To quantitatively illustrate the impact of residual motion artifacts, we measured the area of the segmented lumen and plotted it against the motion amplitudes. The thresholding-based segmentation for the lumen and necrotic core is presented in Fig.~\ref{fig:MotionAnalysis}(a) with the curve of the lumen area measurement with respect to various residual motion amplitudes, as shown in Fig.~\ref{fig:MotionAnalysis}(b). Interestingly, the segmentation results are more sensitive to the cardiac motion compared to their image visual appearances in Fig.~\ref{fig:MotionAnalysis}(a). The motion artifacts at the 20\% level in Fig.~\ref{fig:Motion} look negligible while the shape of the segmented vessel wall and necrotic core already changes noticeably from the corresponding image. However, the area measurement of the segmented lumen does not suffer much from the motion artifacts if the residual movement is less than 40\%, as reflected in Fig.~\ref{fig:MotionAnalysis}(b), despite slight positional shifting of the segmentation in Fig.~\ref{fig:MotionAnalysis}(a). This finding implies that the quantitative analysis of a bright spot (e.g., contrast enhanced blood pool, relatively uniform high-contrast calcification) is less easily affected by motion artifacts than a less dense structure (e.g., necrotic core, soft tissue). 
Therefore, to achieve the best image quality for both high- and low-density structural analysis, the residual movement should be controlled below 20\% of the original heart movement.

\subsection{Spectral Characterization}
\begin{figure}
  \centering
  \includegraphics[width=0.9\linewidth]{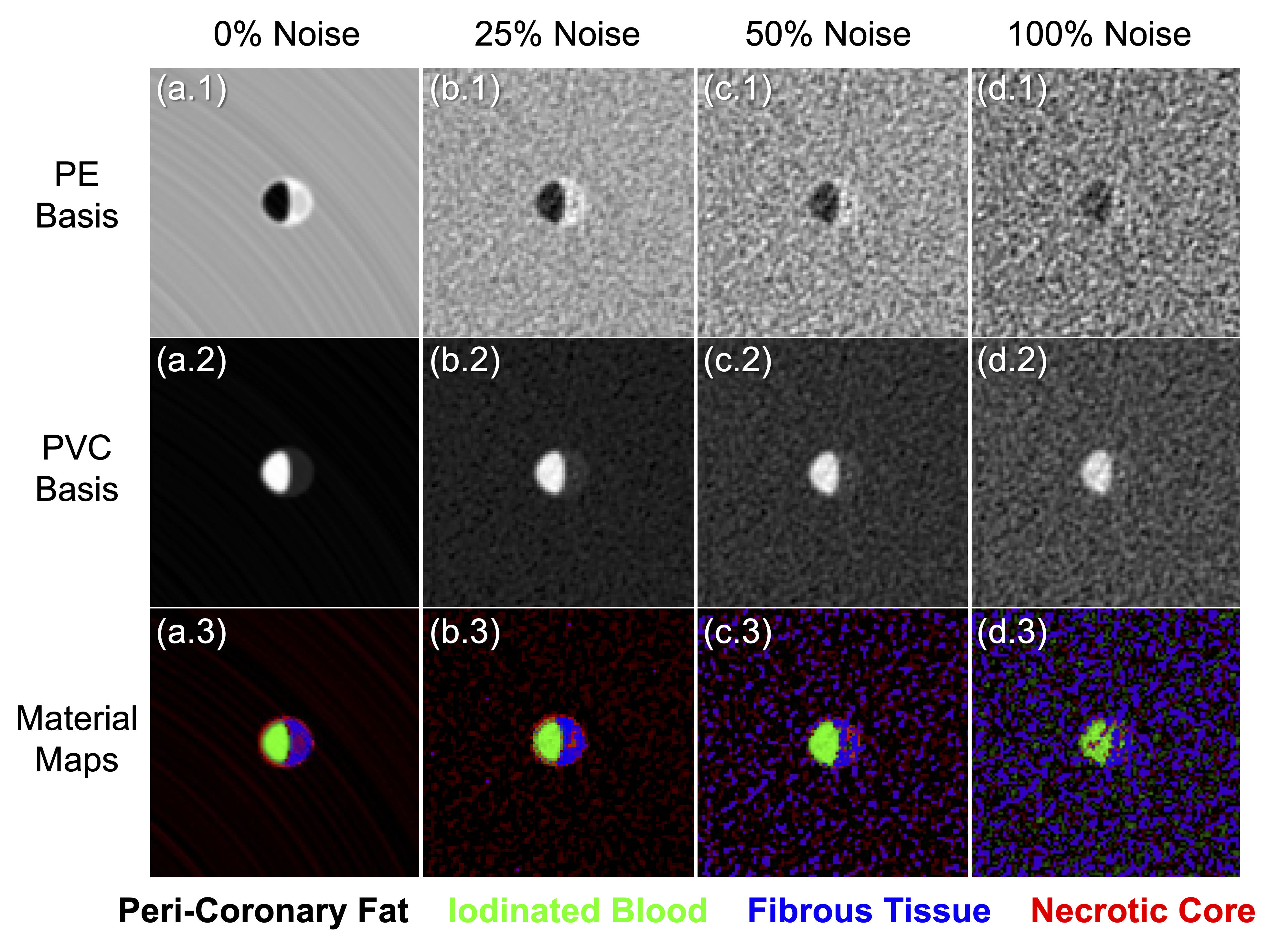}
  \caption{\mz{Basis images (PE and PVC, first and second columns respectively) and material maps (third column) for simulated PCCT plaque images. These images were simulated at different noise level: (a) noise-free case; (b) 25\% of the standard noise level; (c) 50\% of the standard noise level; (d) standard noise level.}}\label{fig:SpectralColor}
\end{figure}

Spectral tissue characterization (i.e., generating material maps from basis images) is one of the major applications of PCCT. Figure~\ref{fig:SpectralColor} illustrates the tissue characterization performance of simulated plaque images at different noise levels. At lower noise level (e.g., 0\% and 25\% of the standard noise), the necrotic core (isolated red “blobs” within the blue fibrous tissue) can be clearly visualized as shown in Fig.~\ref{fig:SpectralColor}(a) and (b). Other materials, including contrast-enhanced blood (green) and fibrous tissue (blue) are also distinguishable from the background peri-coronary fat tissue. Robust identification of the necrotic core is currently not possible at the standard noise level as shown in Fig.~\ref{fig:SpectralColor}(c). Note that these preliminary results were obtained (i) from purely analytical material decomposition methods; and (ii) without pre-processing (e.g., denoising) of the basis images. Performance of both aspects can be greatly improved with data driven methods and are subjects of our ongoing work.

\subsection{Radiation Dose Control}
\begin{figure}
  \centering
  \includegraphics[width=0.8\linewidth]{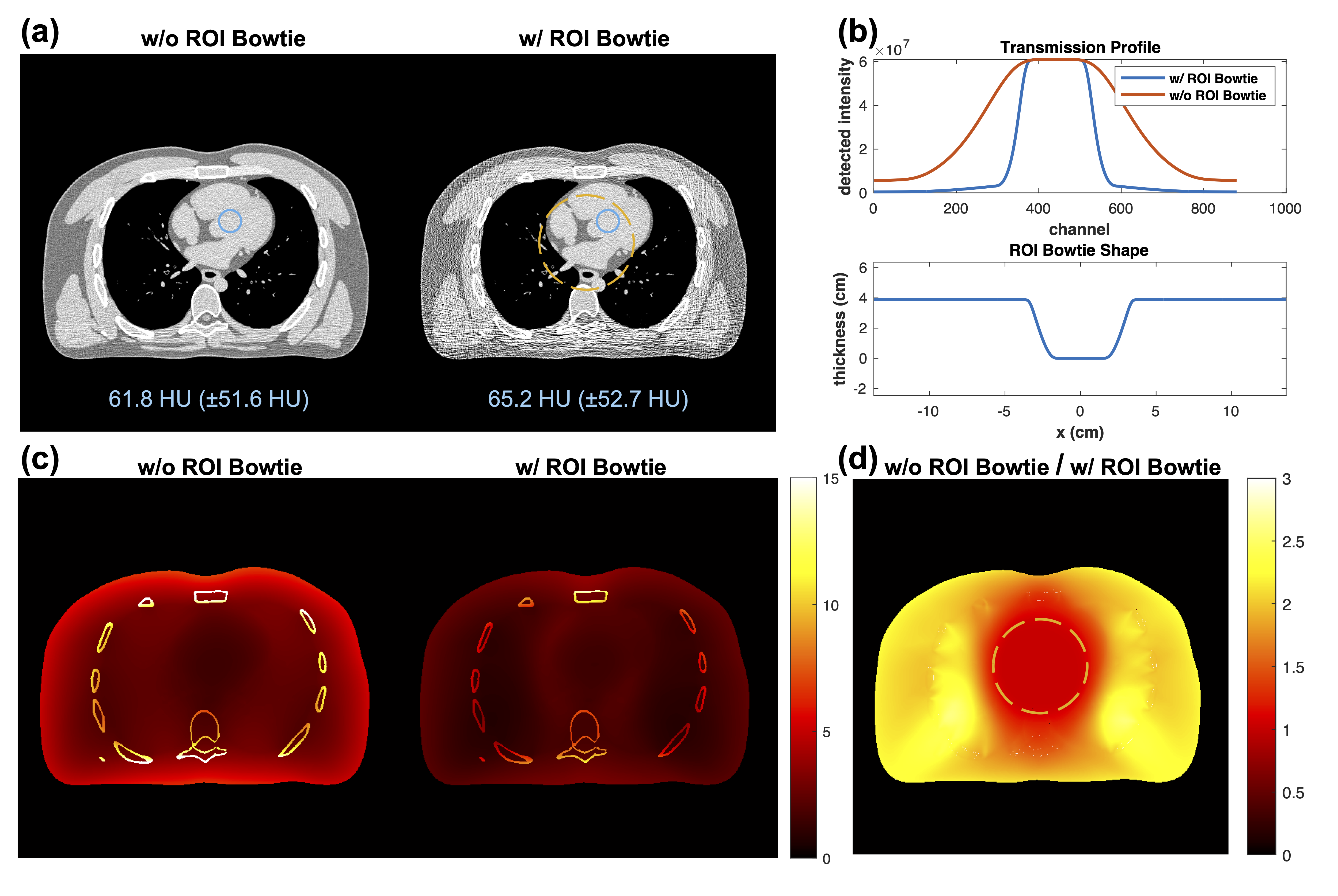}
  \caption{Effects of the ROI bowtie on image noise and radiation dose simulated with a realistic human thorax slice, compared against with only the large bowtie. (a) Noise demonstration shows the ROI bowtie help maintain the noise level within the ROI (circled in orange) while magnifying noise in the peripheral region, and the blue smaller circles mark the region for quantitative noise comparison (the mean and standard deviation values appear beneath the images in blue). The reconstruction images are displayed in the same window (C-100 HU/W500 HU). (b) The transmission profile through air with the ROI bowtie is displayed on the top demonstrating a huge difference between the X-ray flux within and outside the ROI range while the shape of the ROI bowtie (assuming at a virtual position distanced 26.3~\cm from the iso-center) is shown at the bottom. (c) Absorbed radiation dose reconstructions (with noise turned off) are displayed with a color bar on the right, and (d) the ratio between the two cases without and with the ROI bowtie is also given for comparison.}\label{fig:HumanDose}
\end{figure}

Radiation dose is a major concern for high-resolution medical imaging. A dynamic ROI bowtie can effectively address this issue, such as for cardiac imaging of our interest. Fig.~\ref{fig:HumanDose} illustrates the effectiveness of a such bowtie on limiting the overall dose without increasing the image noise in our ROI. Fig.~\ref{fig:HumanDose}(a) shows the simulated noisy reconstruction of a human thorax slice with and without using an ROI bowtie. Outside the ROI region (delineated with an orange circle), the conventional source filtration without an ROI bowtie presents a relatively clean result while large noise and severe streaks are observed in the case using the ROI bowtie. On the other hand, within the ROI the two results look almost the same as also reflected in the statics over a small flat region (marked with a blue circle) from the heart. The mean values from the two results are 61.83 HU and 65.17 HU respectively, and the standard deviation values are 51.59 HU and 52.68 HU, which are pretty close. The similar standard deviation values suggest comparable noise levels within the ROI between the two cases. Fig.~\ref{fig:HumanDose}(b) shows the transmission profile through air with and without an ROI bowtie in the top row while the bowtie shape is displayed at the bottom. The X-ray flux is heavily dampened outside the ROI covered range as seen from the profile, which helps reduce the radiation deposition outside the ROI and explains the strong noise and artifacts in the peripheral thorax region in Fig.~\ref{fig:HumanDose}(a).  Ideally, the mean and standard deviation values measured in (a) should be the same in both cases as the ROI bowtie does not alter the center transmission profile. The slight differences observed are mainly caused by the different patterns of residual beam hardening artifacts.

To reflect the radiation absorption within the imaging plane, the dose reconstructions under the noise-free condition are shown in Fig.~\ref{fig:HumanDose}(c). The use of ROI bowtie clearly dims the dose map except for the central region (our ROI). For example, the skin, muscles, ribs, and most portions of lungs all appear darker, suggesting less radiation dose experienced. However, around the ROI in the two cases with and without the ROI bowtie share the same dose distribution in both pattern and intensity, which implies the same X-ray flux through the region and could help explain the similar signal-to-noise ratio within ROI shown in Fig.~\ref{fig:HumanDose}(a). This change in the radiation dose map with the ROI bowtie is better visualized as the ratio between the two cases and displayed in Fig.~\ref{fig:HumanDose}; i.e., the result without an ROI bowtie to the result with the ROI bowtie. A value greater than 1 in the figure denotes reduced radiation dose after using the ROI bowtie, and similarly, smaller than unit stands for increased dose. The ratio map demonstrates conspicuous dose reduction on peripheral regions; e.g., over 45\% reduction for side ribs and 55\% reduction for the skin and major muscles. On the other hand, the dose within ROI remains the same as indicated by the flat center disk with almost constant ratio values close to 1. These results demonstrate the improved radiation dose deposition towards our target using an ROI bowtie.

\begin{figure}
  \centering
  \includegraphics[width=0.8\linewidth]{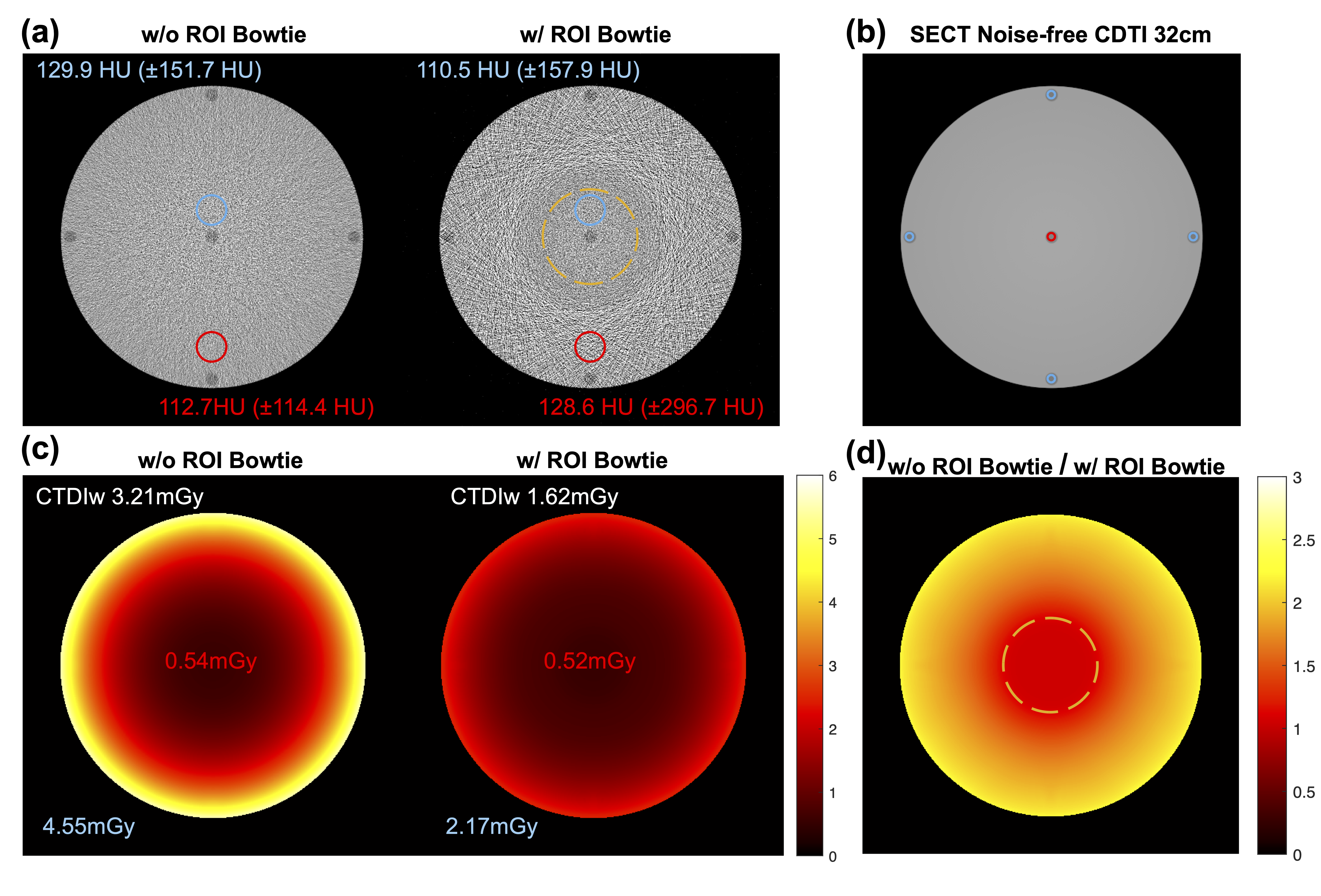}
  \caption{Effects of the ROI bowtie on image noise and radiation dose. (a) Contrary to the case without an ROI bowtie where noise increases from peripheral regions to the center, using the ROI bowtie makes the central ROI have the smallest noise level. The orange circle outlines the ROI while the mean and standard deviation values over regions marked in blue and red are also presented. The display window level and width are C0 HU/W500 HU. (b) The conventional noise-free image of the CDTI phantom is shown with marked regions for \(\text{CTDI}_{w}\) dose index calculation from four peripheral water inserts (blue) and one center water insert (red). (c) Using the ROI bowtie significantly raises radiation dose in the central ROI while lowering that in the peripheral ring. The dose measured from the peripheral and central inserts are added in the figure in the corresponding coding colors, and the calculated \(\text{CTDI}_{w}\) dose values are presented on the top. (d) This dose distribution change is more clearly demonstrated in the ratio map.}\label{fig:CTDIdose}
\end{figure}

More quantitative results are obtained in the CTDI phantom experiments as shown in Fig.~\ref{fig:CTDIdose}. Fig.~\ref{fig:CTDIdose}(a) presents reconstructed noisy images for the CTDI phantom with and without the ROI bowtie. The result with a conventional bowtie demonstrate relatively uniform noise across the whole cross-section although a slight ascending trend of noise levels is still noticeable from outer ring to the inner part. The result with the ROI bowtie suggests a totally different noise level trend; e.g., the noise increases from the out-most ring to the middle ring, then decreases towards the center, and reaches a uniform level within the ROI delineated with an orange circle. The boundary of the center region with uniform noise is actually not clear according to noise texture, and this could be attributed to the gradual slope of the transmission profile as shown in Fig.~\ref{fig:HumanDose}(b). Two smaller regions with one within the ROI (marked with a blue circle) and one outside the ROI (marked with a red circle) are used for noise quantification. The blue and red regions in the conventional bowtie result have mean values of 129.87 HU and 112.72 HU respectively, and corresponding standard deviation values are 151.74 HU and 114.40 HU. The mean values for the result with the ROI bowtie are 110.52 HU and 128.60 HU, and the standard deviations are 157.89 HU and 296.73 HU. The discrepancy between the mean values could result from beam hardening correction. To our interest, the noise level for the blue region within ROI remains similar for two cases, while an increase by 160\% is observed in the peripheral region marked red.

To calculate the dose index in terms of \(\text{CTDI}_{w}\), small regions are selected within five water inserts in the CTDI phantom. The regions are marked on a noise-free image of the phantom without the ROI bowtie as shown in Fig.~\ref{fig:CTDIdose}(b) with those from four peripheral inserts marked with blue circles and the one from the center insert marked with a red circle. The noise-less radiation dose reconstructions are in Fig.~\ref{fig:CTDIdose}(c) and accompanied with measured peripheral dose value (blue text) and center dose value (red text). Due to the phantom symmetry, both dose maps are composed of consecutive concentric rings with their amplitudes decreasing from the outermost to the innermost. Compared to the result without using the ROI bowtie, the result using the ROI bowtie has a less steep slope and a smaller maximum value at the outermost region, resulting a 52\% reduction in peripheral dose (4.55 \(mGy\) against 2.17 \(mGy\)) while maintaining a similar center dose level (0.54 \(mGy\) against 0.52 \(mGy\)). The overall \(\text{CTDI}_{w}\) is calculated as the sum of 2/3 of the peripheral dose and 1/3 of the central dose, showing a 50\% reduction with the ROI bowtie (3.21 \(mGy\) against 1.62 \(mGy\)). The changing trend of the dose map is more clearly visualized in Fig.~\ref{fig:CTDIdose}(d) as a ratio map, which reflects a gradual transition from a dose reduction to a dose neutral plateau from an outer layer to the inner part. This transition optimizes the radiation deposition for our goal of cardiac imaging. 

In a brief summary, with the ROI bowtie, we have maintained the noise level within our 10~\cm diameter ROI at only 50\% of conventional overall dose (\(\text{CTDI}_{w}\)). As the ROI bowtie trims the beam width of free-passing X-rays and the radiation dose depends on the size of the ROI, a higher dose reduction can be expected for a smaller desired ROI.

\section{Discussions}
Plaque characterization is a holy grail of CT imaging, and there are extensive studies reporting CCTA imaging with traditional SECT~\cite{motoyama2007atherosclerotic,saremi2015coronary,becker2003ex} or more recent Dual energy CT~\cite{barreto2008potential,vliegenthart2012dual,ding2021characterization}. 
While contemporary CCTA results with the latest PCCT technology are all based on PCDs made of CZT/CdTe material, this study mainly focuses on characterizing plaque imaging with realistic simulation of a deep-silicon PCCT prototype, which stands for an alternative photon-counting CT technology and will play an important role in the future. Due to their distinct design principles, CZT/CdTe-based PCCT and silicon-based PCCT are believed to have different imaging characteristics~\cite{danielsson2021photon}. A side-by-side comparison between the two approaches should be an interesting topic for further research.

There are also several limitations in this simulation study. In the spatial resolution study, although we have demonstrated the great visibility of necrotic core with PCCT imaging which has been challenging to see with conventional EID-CT, the spatial resolution of deep-silicon PCCT is still insufficient to directly image and measure a thin cap lid less than 65~\(\mu m\) for direct classification of vulnerable plaque. Nevertheless, the improvements in spatial resolution and tissue contrast are already beneficial enough to substantially improve current plaque characterization with CT. Additionally, the visibility of a necrotic core is related to its size, and a smaller necrotic core may requires a more aggressive resolution to detect. On the other hand, a tiny necrotic core is much less danger than a bigger one since ``a plaque with a core less than 3~\mm, with an area of less than 1.0~\(mm^2\)'' is thought unlikely to rupture~\cite{saremi2015coronary}. 

In the noise influence study, current reconstructions used the analytic filtered back-projection algorithm while in practice PCCT images are often reconstructed with more advanced iterative reconstruction methods or even deep learning algorithms. Since different reconstruction techniques could alter the noise characteristics, it is possible that we could reach a different optimal combination of voxel size and reconstruction parameters when applying those more advanced reconstruction techniques. Rather than providing a direct solution with the best reconstruction setting, the most important point of this study is to demonstrate the importance of selecting an appropriate reconstruction setting and also illustrate a method to search for the optimal settings which could be applied in practice. Additionally, although we try to make the simulation realistic, there may still be gaps between simulation results and real-world practice. How well the noise in this simulation study align with real-world experience remains an interesting question to discuss and will the future work.

In the motion artifact study, our different levels of residual motion artifacts are achieved through simple scaling for all views while in practice residual motion levels could be different among views, and they may not follow the sinusoidal pattern either, which could affect the result. Such effects can be studied in the future. Clearly, higher resolution PCCT demands higher standards in motion correction and more advanced methods with greater accuracy. A similar point has also been demonstrated in our previous work~\cite{li2022motion,li2022motion2}.

In the radiation dose study, the ROI bowtie used in simulation is currently static, which could meet our need for cardiac imaging by centering the heart at the iso-center of the imaging plane. But for the most dose efficiency way, a dynamic bowtie should be used for a patient centered scan~\cite{chen2009dual}. The dose difference between a heart centered scan and a patient centered scan may be very small due to small position offsets between the two scan modes as implied in Fig.~\ref{fig:HumanDose}. But for other applications where the ROI is not close to the patient center, the difference could be significant, and a dynamic bowtie would become necessary, which will involve complex hardware modifications. 

\section{Conclusion}
In this work, we have demonstrated the feasibility of plaque characterization with deep-silicon PCCT through simulation-based studies in the aspects of spatial resolution, noise influence, motion artifacts, spectral characterization, and radiation dose control. A few interesting points are summarized as follows:
\begin{enumerate}
  \item We have built a voxelized 3D digital plaque phantom with clinically meaningful geometry and realistic chemical composition, and used different voxel sizes for the foreground and the background representation for computational efficiency, making it a great fit for volumetric simulation studies of high-resolution photon-counting CT imaging.
  \item We have suggested a desired spatial resolution of 0.43~\mm~(FWHM) for a clear visibility of the necrotic core and reliable quantitative measurements in a noise-less and motion-free scenario. 
  \item We have found that noise has a huge influence on the presentation clarity of the necrotic core, and appropriate reconstruction kernel type and voxel size help suppress the noise. For the deep-silicon prototype, we have found the combination of 0.25~\mm~voxels and the soft kernel provides the most visually appealing result with the best detectability of the necrotic core in our numerical experiments.
  \item We have noticed that PCCT has stricter requirements on residual motion correction than that with the EID-CT with a lower resolution. Additionally, quantitative measurements appear more sensitive to the residual motions than the visual quality, and the residual motion amplitude should be controlled below 20\% of the heart movement (\(<\) 0.4~\mm~in residual movement range) for accurate image analysis. 
  \item We have justified the radiation dose of PCCT cardiac imaging to be comparable to that of a regular cardiac scan by incorporating advanced iterative reconstruction and deep learning techniques (5.7x dose reduction), which can be further optimized with an ROI bowtie (8x dose reduction). Our radiation simulation demonstrates that the ROI bowtie optimizes the dose deposition for cardiac imaging and cut the overall radiation dose by half to maintain the same noise level within an ROI of 10~\cm diameter. As evidenced by the literature, deep denoising techniques can help conservatively cut 75\% further. With all these combined, a dose reduction factor more than 8 is expected. Thus, we would expect the ultimate PCCT image quality be better than the noisy high-resolution results with 2.8 seconds exposure time shown in Fig.~\ref{fig:NoiseAndVoxel_noisefree}(c), which was simulated with 8 times longer exposure time without any of these techniques employed.   
\end{enumerate}

In conclusion, the results of this simulation study suggest that the deep-silicon PCCT prototype could provide sufficient spatial resolution and tissue contrast for excellent plaque characterization. Together with advanced reconstruction techniques and/or an ROI bowtie, the radiation dose of a cardiac PCCT scan can be controlled to the same level of a conventional EID-CT scan. The increased sensitivity of PCCT imaging to residual motion artifacts also suggests that more advanced correction methods are needed to reach its full potential. We believe that PCCT will bring us to the new era of cardiac imaging.

\bibliographystyle{spiebib} %

\end{document}